\documentclass[twocolumn]{aastex63}

\begin{document}

\title{Final {\it Spitzer} IRAC Observations of the Rise and Fall of SN 1987A}

\author[0000-0001-8403-8548]{Richard G. Arendt} 
\affiliation{Code 665, NASA/GSFC, 8800 Greenbelt Road, Greenbelt, MD 20771, USA}
\affiliation{CRESST II/UMBC, USA; Richard.G.Arendt@nasa.gov}

\author[0000-0001-8033-1181]{Eli Dwek} 
\affiliation{Code 665, NASA/GSFC, 8800 Greenbelt Road, Greenbelt, MD 20771, USA}

\author[0000-0002-6018-3393]{Patrice Bouchet}
\affiliation{Laboratoire AIM Paris-Saclay, CEA-IRFU/SAp, CNRS, Universit\'e Paris Diderot, F-91191 Gif-sur-Yvette, France}

\author{I. John Danziger}
\affiliation{INAF-Osservatorio Astronomico di Trieste, via G.B. Tiepolo 11, 34143 Trieste, Italy}

\author[0000-0003-1319-4089]{Robert D. Gehrz}
\affiliation{Minnesota Institute for Astrophysics, School of Physics and Astronomy,
University of Minnesota,
116 Church Street, SE, Minneapolis, MN 55455, USA}

\author[0000-0003-3900-7739]{Sangwook Park}
\affiliation{Department of Physics, University of Texas at Arlington, Arlington, TX 76019, USA}

\author[0000-0001-6567-627X]{Charles E. Woodward}
\affiliation{Minnesota Institute for Astrophysics, School of Physics and Astronomy,
University of Minnesota,
116 Church Street, SE, Minneapolis, MN 55455, USA}

\begin{abstract}
\indent
{\it Spitzer}'s final Infrared Array Camera (IRAC) observations of 
\object{SN 1987A} show the 3.6 and 4.5~$\mu$m 
emission from the equatorial ring (ER) continues a period of steady decline.
Deconvolution of the images reveals that the emission is dominated by the 
ring, not the ejecta, and is brightest on the west side.
Decomposition of the marginally resolved emission also confirms this,
and shows that the west side of the ER has been brightening 
relative to the other portions of the ER. The infrared (IR) morphological 
changes resemble those seen in both the soft X-ray 
emission and the optical emission.
The integrated ER light curves at 3.6 and 4.5~$\mu$m are more similar to the 
optical light curves than the soft X-ray light curve, though differences would be expected 
if dust is responsible for this emission and its destruction is rapid. 
Future observations with the {\it James Webb Space Telescope} 
will continue to monitor the ER evolution, and 
will reveal the true spectrum and nature of the material responsible for
the broadband emission at 3.6 and 4.5 $\mu$m. 
The present observations also serendipitously reveal 
a nearby variable source, 
subsequently identified as a Be star, that 
has gone through a multi-year outburst during the course of these observations.
\end{abstract}


\section{Introduction} \label{sec:intro}

The {\it Spitzer Space Telescope} \citep{Werner:2004,Gehrz:2007} was launched 
more than 16 years after the explosion of supernova (SN) 1987A. While far too 
late to observe the explosion itself, the timing of the {\it Spitzer}
mission has been ideal for observing the subsequent interaction 
of the SN blast wave with its circumstellar medium (CSM) \citep{McCray:2016}. That interaction 
had been anticipated ever since it was realized that there was a 
dense, structured CSM surrounding the progenitor star 
\citep{Fransson:1989,luo:1994}. This CSM was 
ionized by the flash of the SN explosion and faded thereafter as the 
gas recombined \citep{Lundqvist:1991,Dwek:1992b}. High resolution ground-based and
{\it Hubble} (Faint Object Camera) images revealed that the 
CSM was dominated by an equatorial ring (ER) and two larger fainter
rings displaced in the poleward directions \citep{Crotts:1989,Wampler:1990,Jakobsen:1991}. 
In 1995, the first hotspot in the ER appeared \citep{Pun:1997,Lawrence:2000} 
as the fastest SN ejecta 
began impacting the innermost portion of the ER. Other hotspots appeared and brightened 
from 1999 through 2009, when the ER was fully delineated by 
dozens of hot spots \citep{Bouchet:2000,Fransson:2015}. 
As the blast wave swept further through and 
past the ER, the hotspots are now fading and new (though fainter) 
structures are being illuminated beyond the ER, but not yet out to the
polar rings. \citep{Fransson:2015}.

As summarized by \cite{Frank:2016}, the X-ray emission from 
SN~1987A provides a complementary view of the interaction of the 
SN with its surrounding CSM. Following the initial detection of 
the early hot spots, the onset of the main interaction with the 
ER in 2003 (Day $\sim6000$) was marked by a distinct decrease 
in the rate of expansion of the X-ray emitting gas and a sharp 
rise in the soft X-ray emission. Continued twice-yearly monitoring 
with {\it Chandra} has documented the rising X-ray emission from
the ER through 2014 (Day $\sim10000$), and the development of an 
asymmetry brightening of the western ER emission at the later times 
(after Day $\sim7500$).

{\it Spitzer} has provided unique mid-IR observational capabilities for studying 
the interaction of SN 1987A with the ER. Spectroscopy with the InfraRed Spectrograph (IRS) instrument \citep{Houck:2004} 
at 5 -- 30 $\mu$m revealed that the mid-IR 
is dominated by emission from warm silicate dust at an 
apparently uniform temperature 
of $T_{\rm d} \approx 180$~K \citep{Bouchet:2006}. This confirmed hints
of silicate emission that had been detected by the
{\it Infrared Space Observatory (ISO)} \citep{Fischera:2002}.
Comparison with higher angular
resolution ground-based imaging indicated that the dust was located in the ER.
Subsequent observations showed that the mid-IR spectrum brightened, but with 
no clear change in the temperature of the dust \citep{Dwek:2008,Dwek:2010}.

Monitoring with the Multiband Imaging Photometer 
for {\it Spitzer} (MIPS) instrument \citep{Rieke:2004} 
with broadband photometry at 24 and 70 $\mu$m, showed consistency 
with the IRS measurements at 24 $\mu$m, but failed to detect any emission
from colder dust at far-IR wavelengths. Colder dust was subsequently detected
with {\it Herschel} and Atacama Large Millimeter/submillimeter Array 
(ALMA), but this colder component is associated 
with the slower moving dense ejecta, well inside of the ER 
\citep{Matsuura:2011,Indebetouw:2014,Cigan:2019}. Recently, 
\cite{Matsuura:2019} detected strengthening emission at 31~$\mu$m 
using the Faint Object InfraRed CAmera for the 
SOFIA Telescope (FORCAST) instrument \citep{Herter:2012} on the
Stratospheric Observatory for Infrared Astronomy (SOFIA) 
\citep{Gehrz:2009, Young:2012}. However, it
is not certain whether this emission arises in the ER, the ejecta, or both.

Broadband imaging at 3.6, 4.5, 5.8, and 8~$\mu$m with {\it Spitzer's}
IRAC instrument \citep{Fazio:2004} also confirm the evolution of the 
ER emission seen with the IRS. However, this shorter wavelength 
spectral energy distribution (SED)
requires another emission component in addition to the 180~K 
silicate dust. The 3.6 and 4.5~$\mu$m photometry extend an 
approximately power-law spectrum seen by IRS at 5-10 $\mu$m 
to shorter wavelengths. \cite{Dwek:2010} modeled this 
emission to determine that it was likely from a hotter, 
$T_{\rm d} > 350$~K, dust component. The lack of any 
distinguishing spectral features made it difficult to determine
the likely composition or even the location of this inferred dust.

After {\it Spitzer's} He cryogen was exhausted in 2009, only the 
IRAC 3.6 and 4.5~$\mu$m bands remained operational at the warmer 
spacecraft temperatures. While the exact nature of the emission 
at these wavelengths has not been certain, regular observations
of SN 1987A continued for the purposes of monitoring the 
evolution of the interaction with the ER and to develop 
a clear picture of the evolving relationship between the IR emission
at these wavelengths and emission at optical and X-ray 
wavelengths \citep{Dwek:2008,Dwek:2010,Arendt:2016}.

This paper reports on the complete set of {\it Spitzer's}
IRAC observations of SN 1987A and its ER. Section 2 
briefly describes the observations and additional 
mosaicking procedures. The complete mid-IR light curves 
from {\it Spitzer} IRAC and MIPS photometry are presented 
and modeled with a simple empirical function in Section 3.
The final light curves extend a little over 4 years 
(in 8 epochs) beyond the results presented by \cite{Arendt:2016}.
In section 4 we employ high resolution mapping and 
deconvolution that are enabled by the long-term repeated 
observations of SN1987A at a wide range of position angles. We
model the marginally resolved ER to identify the 
separate evolution of the N, S, E, and W portions of the ER.
Section 5 provides further discussion of some of the results,
and the findings are summarized in Section 6.
The appendix describes a strongly 
variable source that is unassociated with SN 1987A, but 
was serendipitously located within the field of view of the
IRAC observations.

\section{Data} \label{sec:data}

\subsection{Standard Post-BCD Mosaics and Photometry}
The initial targeted IRAC observations of SN 1987A 
employed 12-second frame times using the small-scale 
Spiral16 dither pattern. Guided by those results,
subsequent SN 1987A observations continued the use of 
12-second frame times, but switched to the 
slightly shallower medium-scale Reuleaux12 dither pattern,
resulting in 125~s of total exposure (10.4~s of signal 
integration per 12-second frame time).
These observations were repeated at roughly 
6-month intervals.
The basic calibrated data (BCD) individual frames were
automatically mosaicked into post-BCD (pBCD) images on $0.6''$
pixel scales. The SN brightness was measured from the pBCD
mosaics using aperture photometry. Because the SN is not well
resolved from nearby stars \citep[Star 2 and Star 3 as designated
by][]{Walker:1990},
the source aperture includes the SN and these stars, and
estimates of their flux density \citep[extrapolated from 
the observations by][]{Walborn:1993} are subtracted from 
the reported result for SN 1987A (Table \ref{tab:fluxes}). 

\begin{deluxetable*}{rcccccrr}
\tabletypesize{\footnotesize}
\tablewidth{0pt}
\tablecaption{SN 1987A Flux Densities}
\tablehead{
\colhead{Day} &
\multicolumn{4}{c}{IRAC} &
\colhead{MIPS} &
& \\
\cline{2-5}
\colhead{Number\tablenotemark{a}} &
\colhead{$S(3.6\ \micron)$} &
\colhead{$S(4.5\ \micron)$} &
\colhead{$S(5.8\ \micron)$} &
\colhead{$S(8\ \micron)$} &
\colhead{$S(24\ \micron)$} &
\colhead{AOR\tablenotemark{b}} & 
\colhead{PID\tablenotemark{c}}
}
\startdata
 6130.09 &    0.58 $\pm$    0.01 &    0.94 $\pm$    0.01 &    1.46 $\pm$    0.02 &    4.60 $\pm$    0.03 & \nodata &  5030912 &    \textbf{124} \\
 6184.08 &     \nodata           &     \nodata           &     \nodata           &     \nodata           &    26.3 $\pm$     1.8 &  5031424 &    \textbf{124} \\
 6487.93 &    0.69 $\pm$    0.01 &    1.25 $\pm$    0.01 &    2.18 $\pm$    0.04 &    6.89 $\pm$    0.04 & \nodata & 11526400 &   3680 \\
 6487.94 &     \nodata           &    1.27 $\pm$    0.02 &     \nodata           &    7.04 $\pm$    0.07 & \nodata & 11191808 &   3578 \\
 6551.91 &     \nodata           &     \nodata           &     \nodata           &     \nodata           &    36.4 $\pm$     1.9 & 11531264 &   3680 \\
 6724.25 &    0.80 $\pm$    0.01 &    1.41 $\pm$    0.01 &    2.34 $\pm$    0.04 &    7.50 $\pm$    0.05 & \nodata & 14357248 &  20203 \\
 6725.68 &    0.77 $\pm$    0.02 &     \nodata           &    2.41 $\pm$    0.05 &     \nodata           & \nodata & 14359040 &  20203 \\
 6734.33 &     \nodata           &     \nodata           &     \nodata           &     \nodata           &    41.7 $\pm$     1.9 & 14381312 &  20203 \\
 6823.54 &    0.88 $\pm$    0.01 &    1.50 $\pm$    0.01 &    2.52 $\pm$    0.04 &    8.01 $\pm$    0.06 & \nodata & 14369792 &  20203 \\
 6824.65 &    0.81 $\pm$    0.01 &    1.52 $\pm$    0.01 &    2.55 $\pm$    0.04 &    8.17 $\pm$    0.05 & \nodata & 14371584 &  20203 \\
 6828.53 &     \nodata           &     \nodata           &     \nodata           &     \nodata           &    44.4 $\pm$     1.9 & 14385408 &  20203 \\
 7156.35 &    0.97 $\pm$    0.01 &    1.77 $\pm$    0.01 &    3.07 $\pm$    0.02 &   10.12 $\pm$    0.04 & \nodata & 17720064 &  \textbf{30067} \\
 7158.88 &     \nodata           &     \nodata           &     \nodata           &     \nodata           &    55.3 $\pm$     1.8 & 17720576 &  \textbf{30067} \\
 7298.80 &    1.06 $\pm$    0.01 &    1.89 $\pm$    0.01 &    3.23 $\pm$    0.02 &   11.04 $\pm$    0.04 & \nodata & 17721344 &  \textbf{30067} \\
 7309.70 &     \nodata           &     \nodata           &     \nodata           &     \nodata           &    59.8 $\pm$     1.9 & 17721856 &  \textbf{30067} \\
 7489.68 &     \nodata           &     \nodata           &     \nodata           &     \nodata           &    65.2 $\pm$     1.9 & 22393600 &  \textbf{40149} \\
 7502.04 &    1.10 $\pm$    0.01 &    2.01 $\pm$    0.01 &    3.59 $\pm$    0.02 &   12.15 $\pm$    0.05 & \nodata & 22393088 &  \textbf{40149} \\
 7687.35 &    1.14 $\pm$    0.01 &    2.10 $\pm$    0.01 &    3.86 $\pm$    0.02 &   12.95 $\pm$    0.04 & \nodata & 22394368 &  \textbf{40149} \\
 7689.55 &     \nodata           &     \nodata           &     \nodata           &     \nodata           &    70.3 $\pm$     1.9 & 22394880 &  \textbf{40149} \\
 7974.80 &    1.18 $\pm$    0.01 &    2.15 $\pm$    0.01 &    3.92 $\pm$    0.02 &   13.52 $\pm$    0.03 & \nodata & 26172672 &  \textbf{50444} \\
 7983.16 &     \nodata           &     \nodata           &     \nodata           &     \nodata           &    75.7 $\pm$     1.9 & 26173184 &  \textbf{50444} \\
 8576.21 &     \nodata           &    2.25 $\pm$    0.02 &     \nodata           &     \nodata           & \nodata & 40242688 &  70020 \\
 8585.63 &    1.22 $\pm$    0.01 &    2.24 $\pm$    0.01 &     \nodata           &     \nodata           & \nodata & 39952896 &  \textbf{70050} \\
 8706.09 &    1.19 $\pm$    0.02 &     \nodata           &     \nodata           &     \nodata           & \nodata & 40245760 &  70020 \\
 8730.61 &    1.22 $\pm$    0.02 &     \nodata           &     \nodata           &     \nodata           & \nodata & 40075008 &  70088 \\
 8732.20 &     \nodata           &    2.25 $\pm$    0.02 &     \nodata           &     \nodata           & \nodata & 40075264 &  70088 \\
 8733.60 &    1.21 $\pm$    0.01 &    2.21 $\pm$    0.02 &     \nodata           &     \nodata           & \nodata & 40075520 &  70088 \\
 8735.25 &     \nodata           &    2.27 $\pm$    0.02 &     \nodata           &     \nodata           & \nodata & 40075776 &  70088 \\
 8736.63 &    1.16 $\pm$    0.02 &     \nodata           &     \nodata           &     \nodata           & \nodata & 40076032 &  70088 \\
 8738.06 &     \nodata           &    2.24 $\pm$    0.01 &     \nodata           &     \nodata           & \nodata & 40076288 &  70088 \\
 8743.47 &    1.27 $\pm$    0.02 &    2.27 $\pm$    0.02 &     \nodata           &     \nodata           & \nodata & 40076544 &  70088 \\
 8751.53 &    1.13 $\pm$    0.02 &    2.16 $\pm$    0.02 &     \nodata           &     \nodata           & \nodata & 40076800 &  70088 \\
 8757.27 &    1.25 $\pm$    0.02 &    2.24 $\pm$    0.01 &     \nodata           &     \nodata           & \nodata & 40077056 &  70088 \\
 8829.32 &     \nodata           &    2.28 $\pm$    0.02 &     \nodata           &     \nodata           & \nodata & 40246784 &  70020 \\
 8856.29 &    1.21 $\pm$    0.01 &    2.24 $\pm$    0.01 &     \nodata           &     \nodata           & \nodata & 39953152 &  \textbf{70050} \\
 9024.97 &    1.19 $\pm$    0.01 &    2.26 $\pm$    0.01 &     \nodata           &     \nodata           & \nodata & 42277120 &  \textbf{80038} \\
 9232.27 &    1.18 $\pm$    0.01 &    2.24 $\pm$    0.01 &     \nodata           &     \nodata           & \nodata & 42277376 &  \textbf{80038} \\
 9495.25 &    1.17 $\pm$    0.01 &    2.15 $\pm$    0.01 &     \nodata           &     \nodata           & \nodata & 47840256 &  \textbf{90117} \\
 9656.20 &    1.18 $\pm$    0.01 &    2.13 $\pm$    0.01 &     \nodata           &     \nodata           & \nodata & 47840512 &  \textbf{90117} \\
 9810.19 &    1.21 $\pm$    0.01 &    2.12 $\pm$    0.01 &     \nodata           &     \nodata           & \nodata & 49253632 &  \textbf{10038} \\
10034.95 &    1.14 $\pm$    0.01 &    2.03 $\pm$    0.01 &     \nodata           &     \nodata           & \nodata & 49253888 &  \textbf{10038} \\
10244.63 &    1.11 $\pm$    0.01 &    1.97 $\pm$    0.01 &     \nodata           &     \nodata           & \nodata & 52540160 &  \textbf{11023} \\
10377.66 &    1.11 $\pm$    0.01 &    1.91 $\pm$    0.01 &     \nodata           &     \nodata           & \nodata & 52540416 &  \textbf{11023} \\
10540.65 &    1.07 $\pm$    0.01 &    1.86 $\pm$    0.01 &     \nodata           &     \nodata           & \nodata & 52540672 &  \textbf{11023} \\
10722.71 &    1.06 $\pm$    0.01 &    1.79 $\pm$    0.01 &     \nodata           &     \nodata           & \nodata & 52540928 &  \textbf{11023} \\
10905.56 &    1.03 $\pm$    0.01 &    1.73 $\pm$    0.01 &     \nodata           &     \nodata           & \nodata & 60648448 &  \textbf{13004} \\
11090.30 &    0.98 $\pm$    0.01 &    1.67 $\pm$    0.01 &     \nodata           &     \nodata           & \nodata & 60648704 &  \textbf{13004} \\
11272.49 &    0.98 $\pm$    0.01 &    1.61 $\pm$    0.01 &     \nodata           &     \nodata           & \nodata & 60648960 &  \textbf{13004} \\
11462.42 &    0.90 $\pm$    0.01 &    1.56 $\pm$    0.01 &     \nodata           &     \nodata           & \nodata & 60649216 &  \textbf{13004} \\
11667.58 &    0.86 $\pm$    0.01 &    1.49 $\pm$    0.01 &     \nodata           &     \nodata           & \nodata & 65861632 &  \textbf{14001} \\
11885.82 &    0.82 $\pm$    0.01 &    1.41 $\pm$ 0.01 &  \nodata         &  \nodata         &   \nodata   & 65861888 & \textbf{14001}    \\
\enddata
\tablecomments{Flux Densities are in mJy. Flux densities 
of [0.41, 0.26, 0.16, 0.09, 0.01] mJy have been subtracted 
at [3.6, 4.5, 5.8, 8, 24] $\mu$m to account for the emission of Stars 2 and 3 in 
the aperture. This correction was not included in Table 1 of \cite{Arendt:2016}.
A machine-readable version of this table is available online.
\label{tab:fluxes}}
\vspace{-2mm}
\tablenotetext{a}{Day number 0 = 1987 Feb 23}
\vspace{-2mm}
\tablenotetext{b}{{\it Spitzer} astronomical observation request (AOR) number}
\vspace{-2mm}
\tablenotetext{c}{{\it Spitzer} program ID number. Bold numbers indicate programs specifically targeting SN 1987A.}
\end{deluxetable*}

Additional survey observations of the Large Magellanic Cloud (LMC) 
have occasionally included SN 1987A as well. Photometry was performed on 
the standard BCD mosaics for these programs also. 
However, the observations are often shallower (collecting 
only 1-3 12-second frame time exposures per AOR, as a quick check 
on variability or with the 
intention that data from several separate coverages 
were to be combined), and thus photometry from the
standard BCD mosaics can be of poorer quality in these cases.

Targeted MIPS 24 $\mu$m observations were obtained at 6 epochs 
during {\it Spitzer's} cryogenic phase. Aperture photometry was 
performed on the standard pBCD MIPS mosaics in a manner similar 
to that used for IRAC. As with IRAC observations, we also include
several incidental observations from survey programs (see Table \ref{tab:fluxes}).

\subsection{Finer Scale Mosaics} 
For the purpose of investigating the structure of SN 1987A
at the highest possible angular resolution, we applied the
self-calibration methods of \cite{Fixsen:2000} and \cite{Arendt:2010}
to the individual BCD frames to correct for residual offset 
effects, and remapped the data onto smaller scale pixels 
using an interlacing method. These mosaics are discussed
in Section \ref{sec:hires}.

\section{Light Curves} \label{sec:light curve}

\begin{figure}[t]
\includegraphics[width=3.25in]{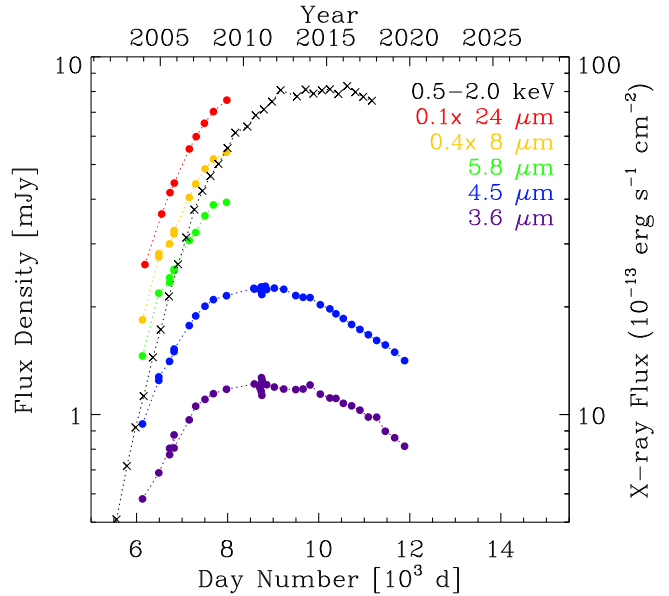}
\caption{SN 1987A light curves throughout the entire {\it Spitzer} mission. 
The {\it Chandra} soft X-ray light curve \citep[0.5-2 keV,][]{Frank:2016} 
is shown for comparison (black $\times$ symbols). \label{fig:light_curve}}
\end{figure}

The multi-wavelength light curves of SN 1987A 
are plotted in Figure \ref{fig:light_curve}. 
With the additional observations, we now see that 
4.5~$\mu$m emission appears to be undergoing a 
very steady decline since Day 9810. This interval of the 
decline is still relatively short, such that it can be fitted
by any of: 
a linear trend (slope $= -0.123 \pm 0.002$ mJy yr$^{-1}$), 
an exponential decay (time scale $= 5210 \pm 60$ days),
or a power law (index $ = -2.08 \pm 0.04$). 

The 3.6~$\mu$m light curve contains more irregularities in its measurements
than the 4.5~$\mu$m light curve. The general trend at 3.6~$\mu$m 
is similar to that at 4.5~$\mu$m,
but the present decline is somewhat slower. However, we caution 
that the 3.6~$\mu$m flux densities are much more sensitive to 
any errors in subtraction of the estimated flux densities of 
Stars 2 and 3, because of the strongly contrasting colors
between the SN and the stars.

\begin{figure*}[t]
   \centering
\gridline{\fig{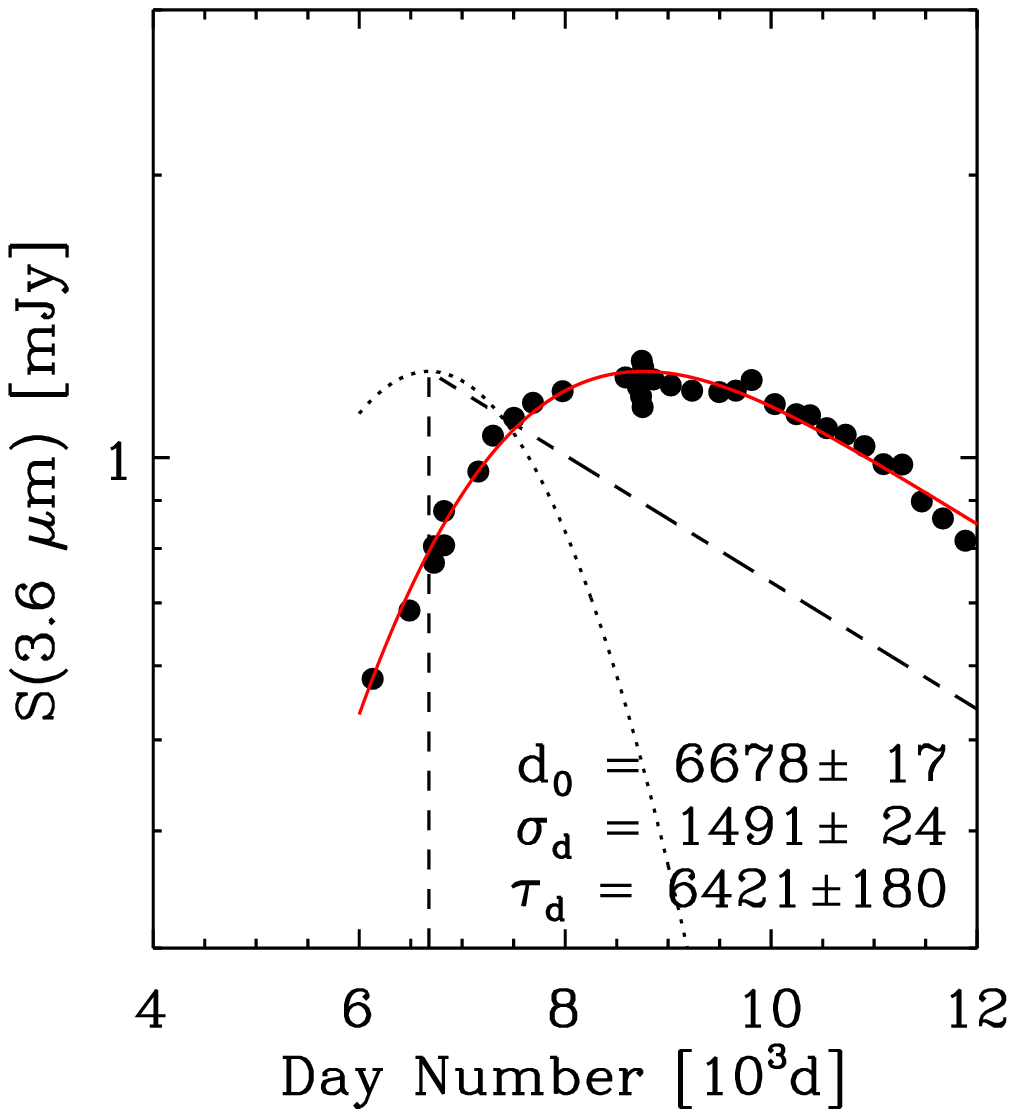}{2in}{}
   \fig{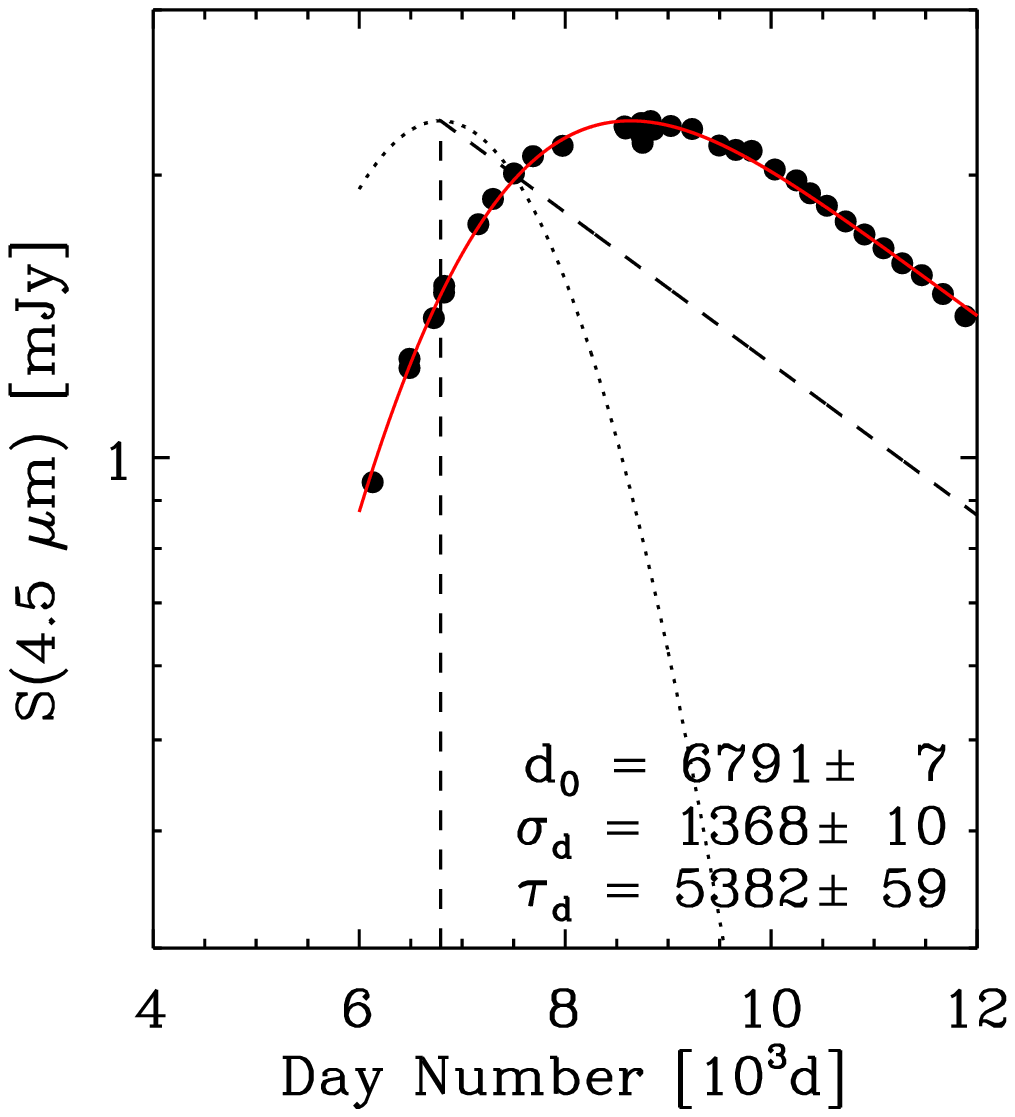}{2in}{}
   \fig{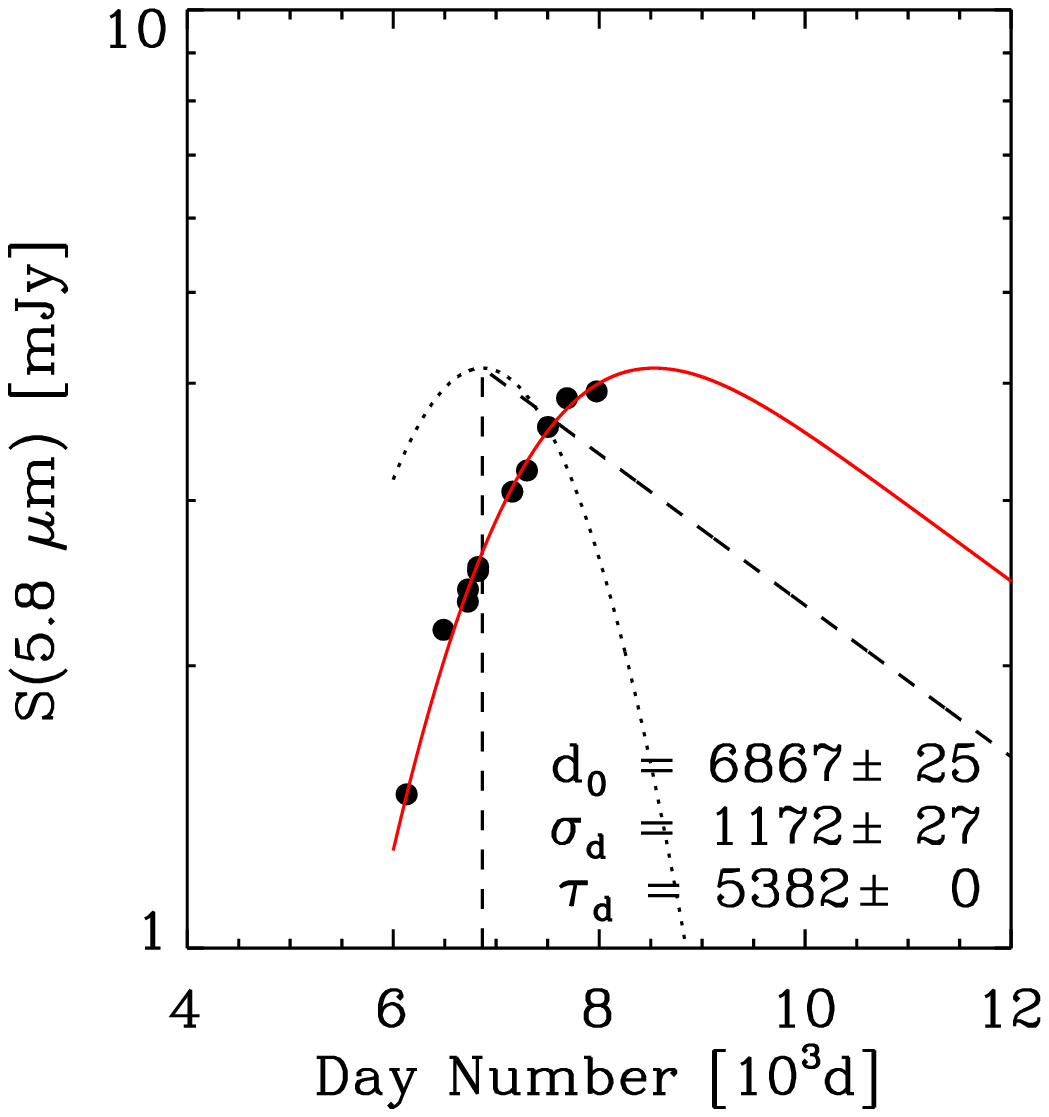}{2in}{}}
\gridline{\fig{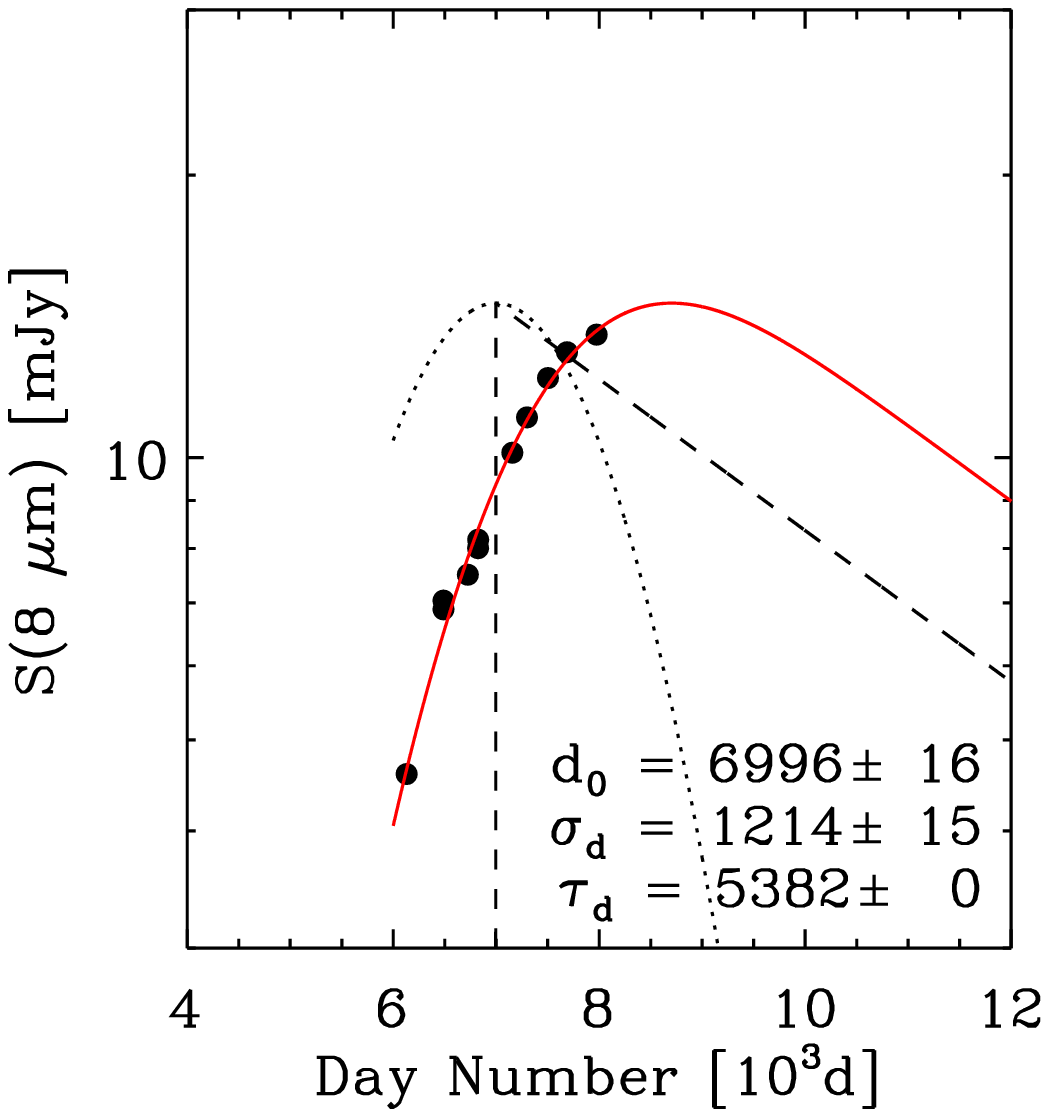}{2in}{}
   \fig{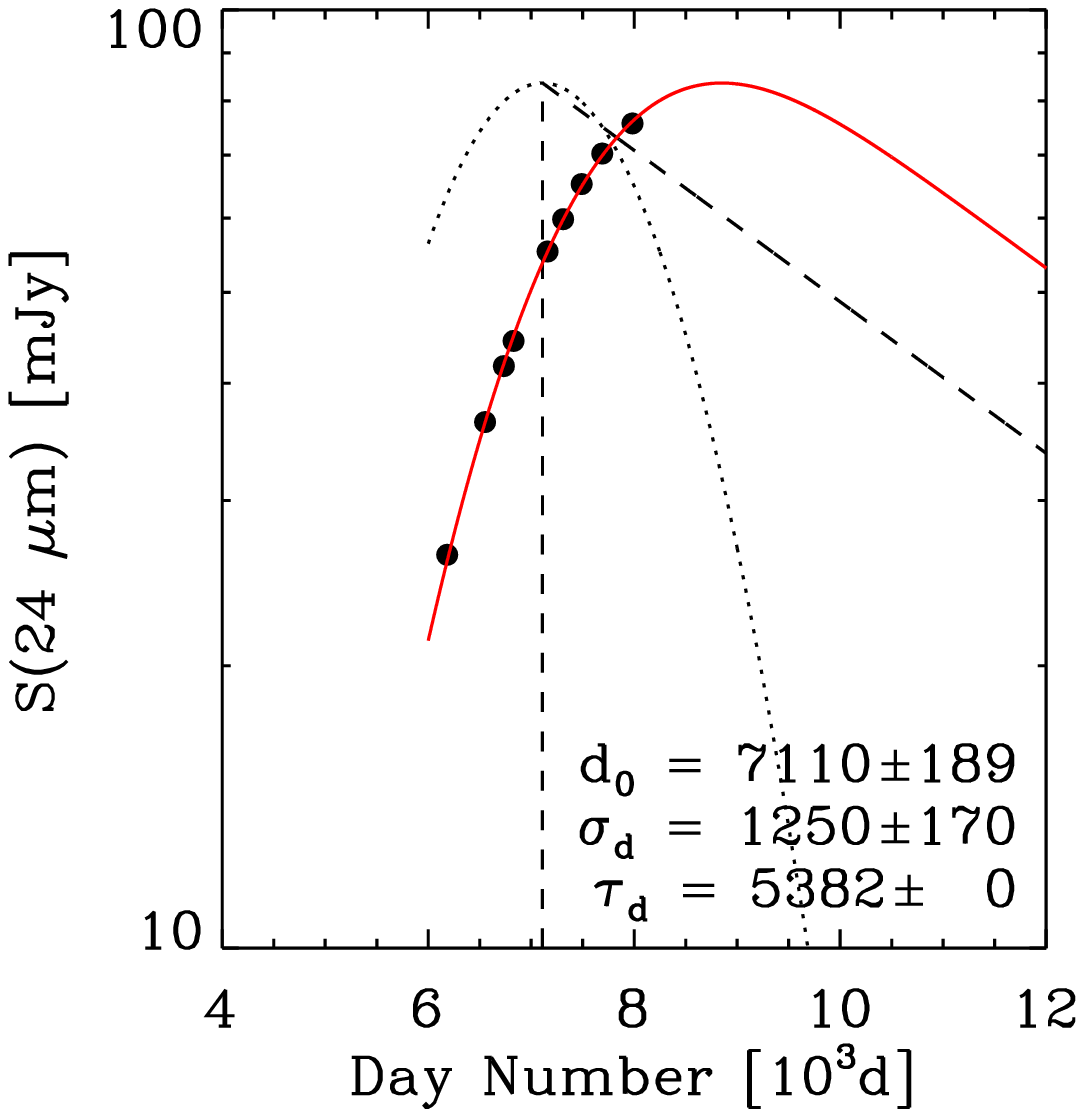}{2in}{}
   \fig{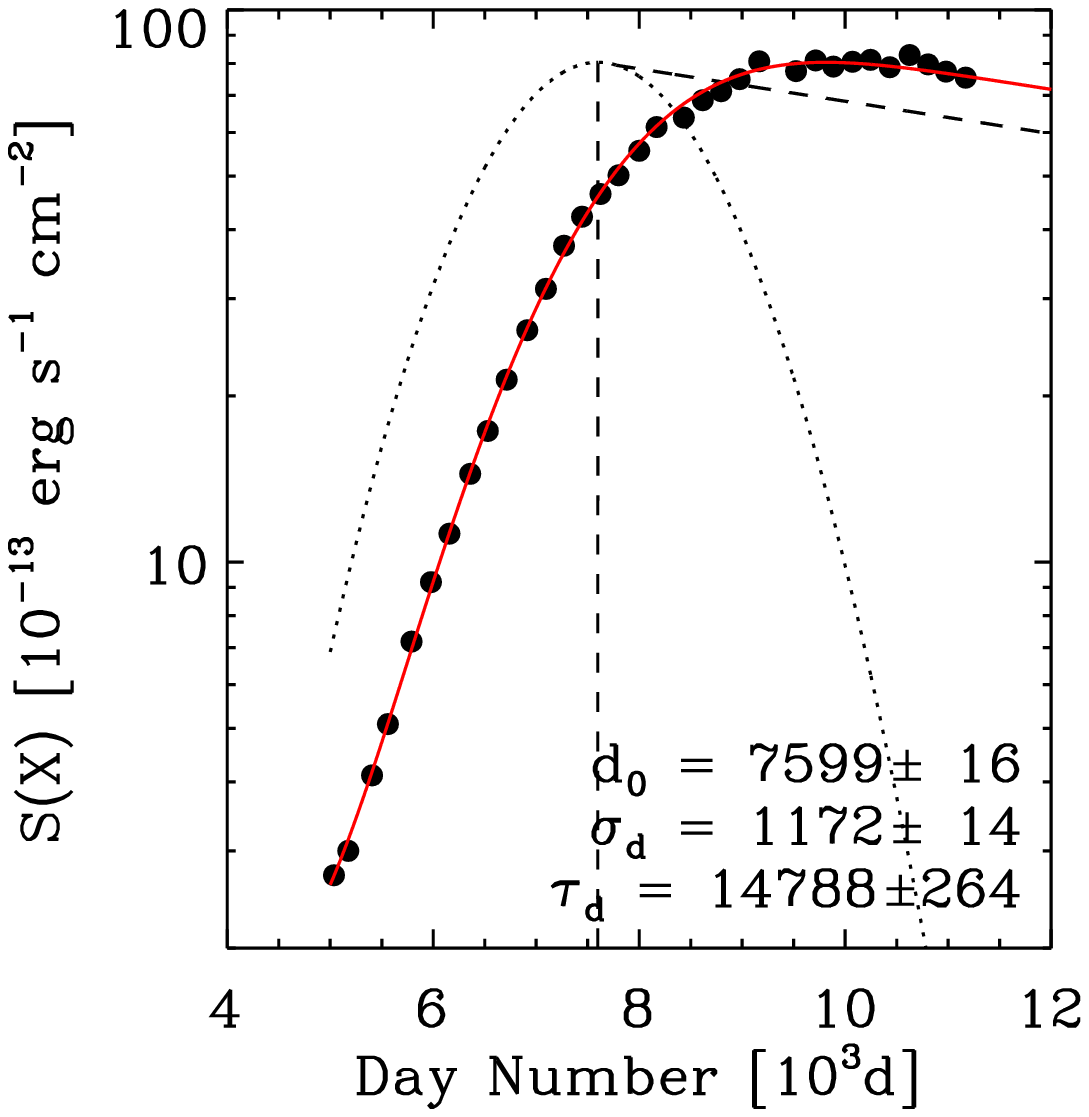}{2in}{}}
\gridline{\fig{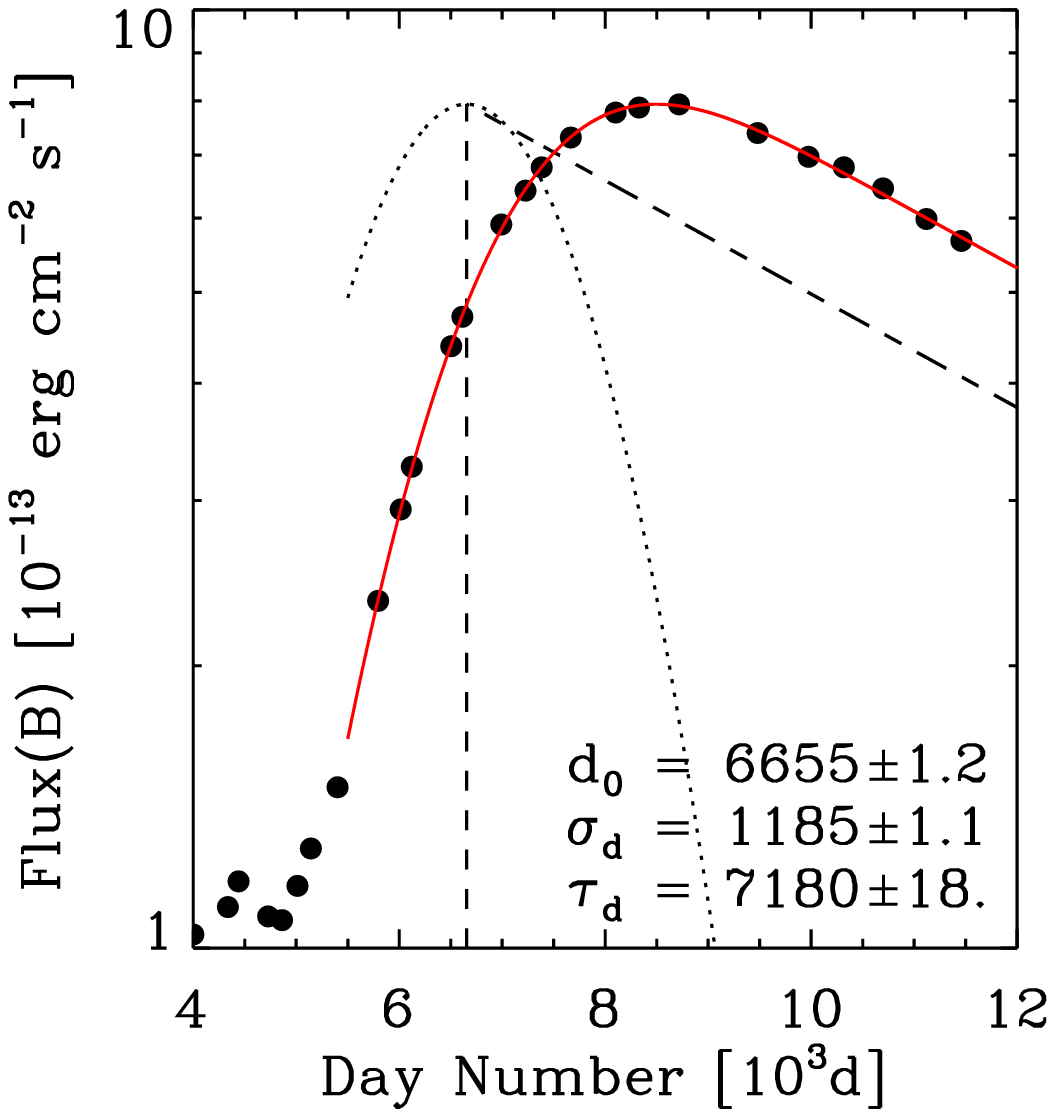}{2in}{}
   \fig{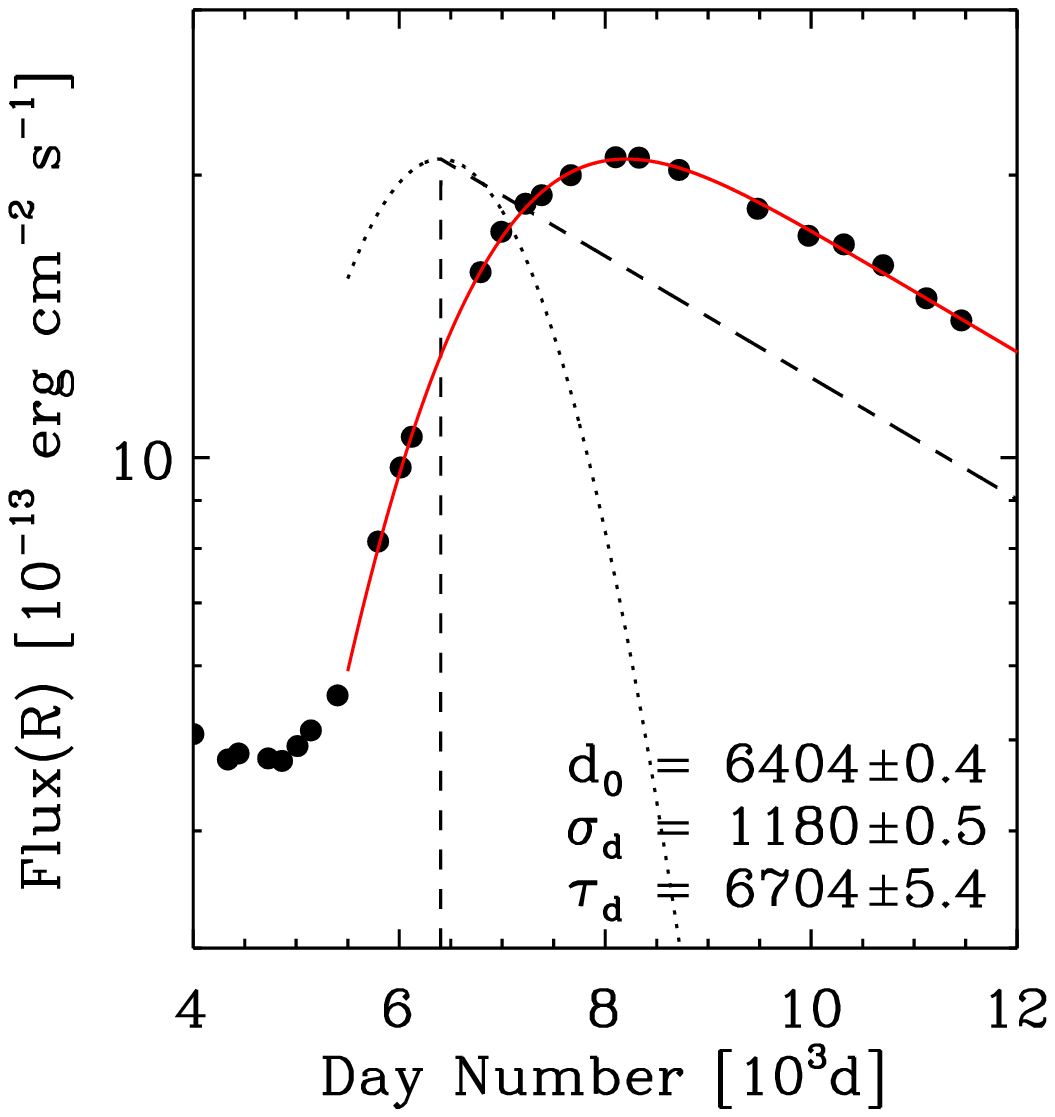}{2in}{}
   \fig{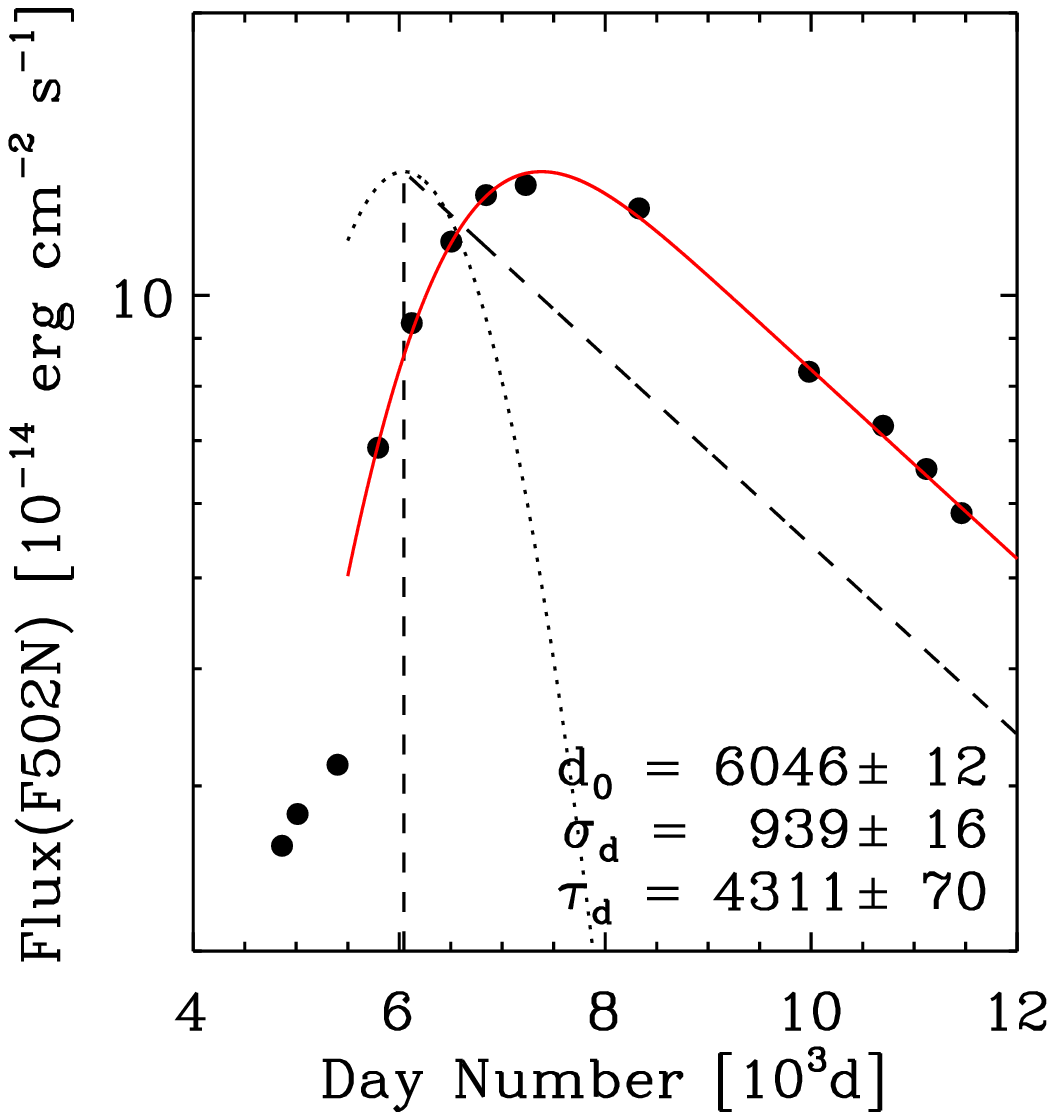}{2in}{}}
   \caption{The {\it Spitzer}, soft X-ray  
   \citep[0.5-2 keV,][]{Frank:2016}, and optical \citep{Larsson:2019}
   light curves (black points) fitted with the model of Equation (1) (red line). 
   For reference, the model components are drawn with dotted and dashed lines 
   with an arbitrary normalization. 
   The center ($d_0$) and width ($\sigma_{\rm d}$) of the gaussian function 
   are listed, as is the time scale ($\tau_{\rm d}$) for the exponential function. 
   At 5.8 -- 24 $\mu$m the parameter $\tau_{\rm d}$ is fixed to the value
   that is derived at 4.5 $\mu$m.
   The intensity range shown for the X-ray emission is wider than 
   that of the other wavelengths.
   \label{fig:model}}
\end{figure*}

Given the apparent exponential decay of the 4.5~$\mu$m light curve,
a model is suggested in which the emission turns on 
as the shock sweeps into the ER, and then progressively decays 
away after the passage of the shock. A very simple form of the 
model would thus be a gaussian function convolved with an exponential 
decay as described by
\begin{eqnarray}
S(t) & = & A \exp[-0.5(t-d_0)^2/\sigma_{\rm d}^2] * \exp(-t/\tau_{\rm d}) \\
     & = & A \int^\infty_{-\infty} \exp[-0.5(t'-d_0)^2/\sigma_{\rm d}^2] \exp[-(t-t')/\tau_{\rm d}] dt' \nonumber
\end{eqnarray}
where $*$ represents convolution.
The free parameters of this model are: an amplitude, $A$, which is 
proportional to the emissivity of the material; a date, $d_0$, which sets a
nominal date for the peak of the interaction; a time scale for 
the gaussian function, $\sigma_{\rm d}$, which sets the scale for 
the rise in the emission; and a time scale for 
the exponential function, $\tau_{\rm d}$, which sets the scale for 
the fading.
Figure \ref{fig:model} depicts the application of this model 
to the {\it Spitzer} data, {\it Chandra} X-ray data \citep{Frank:2016},
and {\it Hubble} optical data \citep{Larsson:2019}.
The X-ray data are supplemented with 4 additional epochs of observations 
(from GO cycles 17 and 18) which were similarly acquired and reduced
as described by \cite{Frank:2016}.
At 5.8, 8, and 24 $\mu$m the decay time scale, $\tau_{\rm d}$, was 
fixed at the value determined at 4.5 $\mu$m 
because of the lack of late time data.

\section{High-Resolution Images: Deconvolution and Decomposition}\label{sec:hires}

\subsection{Deconvolution}
\begin{figure*}[t]
\begin{center}
\includegraphics[width=7in]{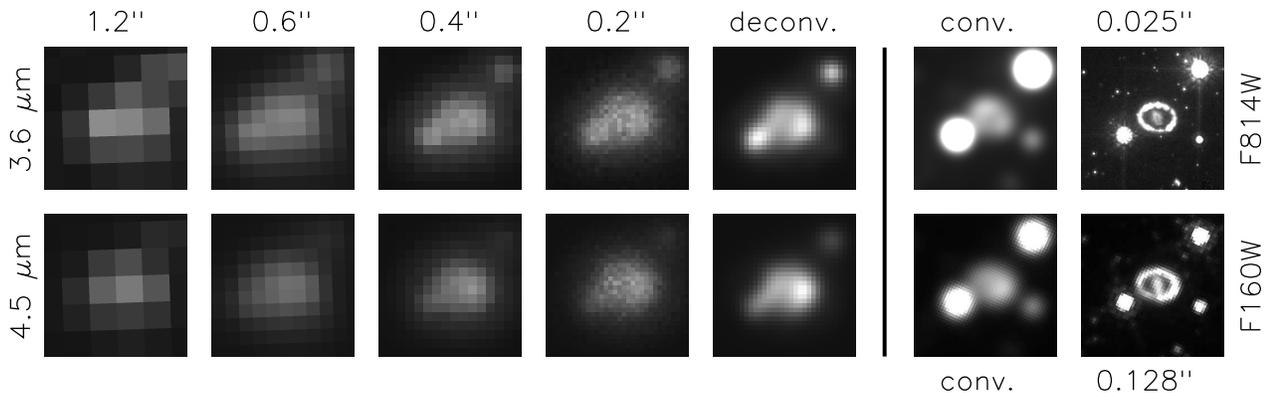}
\caption{High resolution IRAC imaging the SN 1987A ER. The first column 
illustrates the SN 
as seen in a single BCD frame at the intrinsic $1.2''$ pixel scale of the detectors. The second 
column shows a standard pBCD mosaic (generated on $0.6''$ pixels) from the observations at 
one epoch. The third column shows a custom mosaic on $0.4''$ pixels generated from data 
at all epochs. The fourth column is another custom mosaic on $0.2''$ pixels, again using data from all epochs. 
The mosaic with $0.2''$ pixels is then deconvolved to provide the images in the 5th 
column. The appearance of the ER in these images is similar to that in {\it Hubble} 
observations (seventh column) if those observation are convolved to comparable resolution (sixth column).
North is up, and east is to the left. The field of view is $6.4''\times6.4''$.
\label{fig:deconvolve}}
\end{center}
\end{figure*}

The frequently repeated observations of SN 1987A allow mapping the data
into finer scale mosaics than the standard pBCD data product. 
Using data from throughout the mission we have created mosaics with 
$0.4''$ pixel spatial scales
\citep[as in][]{Arendt:2016} and $0.2''$ pixel spatial scales
(Figure \ref{fig:deconvolve}). With $0.4''$ pixels, the ER appears 
better separated from Stars 2 and 3. On $0.2''$ pixels the images
become noisier rather than sharper. The increased noise is because 
the interlacing mapping method leads to very shallow coverage per 
pixel at this scale. The lack of improved resolution is because 
the pixels are already sufficiently small that they no longer 
represent a significant extra convolution of the intrinsic 
point spread function (PSF) in the final mosaics. 

However the $0.2''$ pixel mosaics are suitable for application of
deconvolution techniques. In this study we have processed 
these mosaics using the IDLASTRO procedure {\tt Max\_Likelihood} 
\citep{Landsman:1993}, using 
PSFs generated from the Fourier transform of the idealized {\it Spitzer} 
aperture \citep[outer diameter = 0.85~m, inner 
diameter = 0.32~m,][]{Werner:2004}.
A standard IRAC PSF was not used because
the data were taken at many different position angles, making the effective 
PSF azimuthally symmetric. The diffraction features of the three secondary 
mirror supports are washed out in the effective PSF, and thus omitted in the 
model PSF. The deconvolved $0.2''$ images are shown in Figure \ref{fig:deconvolve}.
In the deconvolved images (especially at 3.6~$\mu$m), the SN is distinct
from stars 2 and 3, and the morphology of the emission is seen to 
be dominated by the ER. The ER appears brighter in the southwest and fainter 
in the southeast. For comparison, the figure also shows that when convolved 
to the same resolution, high-resolution {\it Hubble} images look similar to 
the deconvolved IRAC images. The F814W ACS image was taken at Day 6505,
near the start of the IRAC observations. The F160W WFC3 image was taken at Day 
8718, near the time when the brightness of the ER peaked at 3.6 and 4.5 $\mu$m,
and thus it is more similar to the deconvolved image generated
from the entire span of the IRAC data.
At the shorter wavelengths observed by 
{\it Hubble}, the stars are much brighter relative to the ER.

We find that the deconvolution procedure is insensitive to artificially 
added noise (at a $1\sigma$ level), but 
is mildly sensitive to the PSF, producing spurious results
if attempted with PSFs that are 50\% larger or smaller. The procedure was tried 
on mosaics with larger ($0.3''$ and $0.4''$) pixels, but performed less well.
We used 10 iterations of the procedure, as there was relatively little 
improvement in sharpness with more iterations. We found a 
nearby isolated point source can be fit by a gaussian with full width 
at half maximum (FWHM) of $1.01''$ at 
both 3.6 and 4.5~$\mu$m. Overall, the effectiveness of deconvolution 
achieved here is similar to that demonstrated by \cite{Velusamy:2008}.

\subsection{Decomposition}
\begin{figure}[t]
\begin{center}
\includegraphics[width=3.25in]{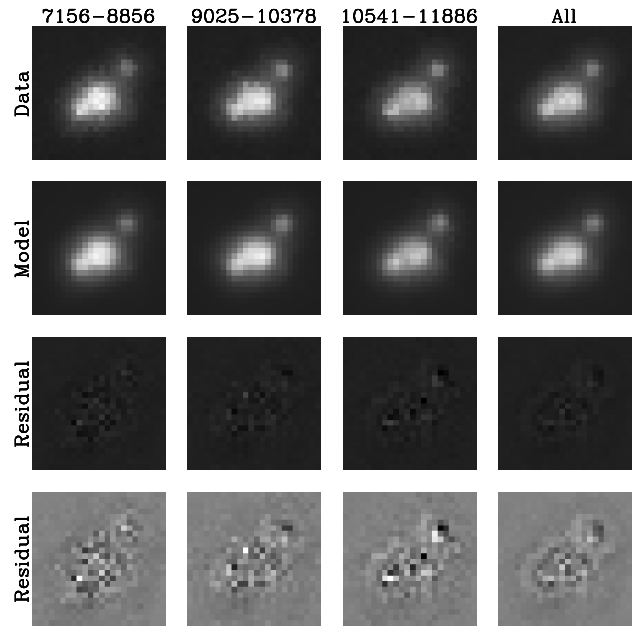}
\caption{(top row) 3.6 $\mu$m images constructed on 
$0.4''$ pixels at different epochs as 
indicated by the day number ranges for each figure. The ``All'' image includes 
all epochs, and is the same as in Figure \ref{fig:deconvolve}.
With increasing time, the ER emission becomes more asymmetric with the southwest 
region becoming the brightest portion at late times. 
Linear scaling from [-1,7] MJy~sr$^{-1}$.
(second row) The corresponding modeled emission at each epoch, displayed on the 
same scale as the images above. (third row) Difference between data and model,
on the same scale. (fourth row) Difference between data and model,
displayed with linear scaling from [-1,1] MJy~sr$^{-1}$.
North is up, and east is to the left. The field of view 
of each panel is $9.6''\times9.6''$).
\label{fig:epochs}}
\end{center}
\end{figure}

We looked for temporal changes in the ER structure by dividing the data set 
into three intervals: early, middle, and late epochs.
The intervals exclude the earliest data when the emission was 
rising rapidly, each contain a similar number of observations, and 
roughly correspond to: the rise to the peak, peak brightness, and decline.
When divided this way, the depth of 
coverage is too low to allow mapping onto $0.2''$ pixels and deconvolution. 
However, mapping on $0.4''$ pixel still works well. Images constructed 
from the early, middle and late epochs are shown in Figure \ref{fig:epochs}.
Despite the limited resolution, comparison of the images indicates that 
the emission was relatively uniformly distributed around the ER at 
the early epochs. At the middle epochs the ER becomes more asymmetric, 
being brighter in the southwest. At late epochs the asymmetry becomes more 
distinct as the southwest fades more slowly than the rest of the ER.

\begin{figure}[t]
\begin{center}
\includegraphics[width=1.65in]{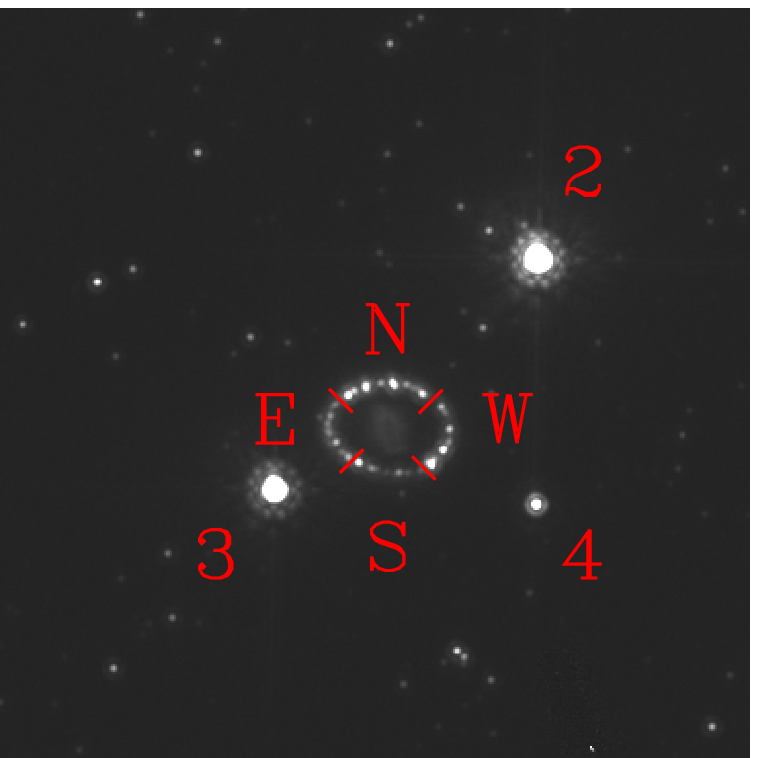}
\includegraphics[width=1.65in]{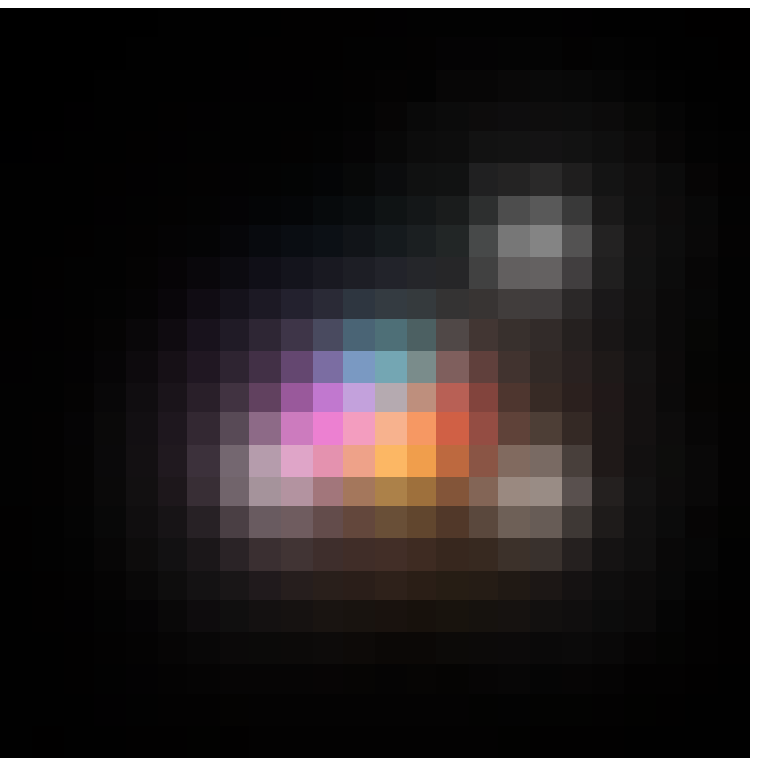}
\caption{The {\it Hubble} image (F814W ACS from Day 6505)
in the top panel is used to 
define the 3.6 $\mu$m emission model's seven components: 
Stars 2, 3, and 4, and four segments of the ER. 
The bottom panel shows these components after convolution with 
the effective 3.6 $\mu$m PSF and mapped on $0.4''$ pixels. For illustration 
purposes, the components are all normalized to unity and the ER components have 
been colorized with 4 different colors. The field of view matches
that used in Figure~\ref{fig:epochs}.
\label{fig:segments}}
\end{center}
\end{figure}

We decompose these $0.4''$ images by modeling the emission as 
the linear combination of 7 components. The first three model
components are Stars 2, 3, and 4 \citep{Walker:1990}.
The ER could be represented by several dozen individual knots; however 
independently solving for the brightness of each knot would lead to 
degeneracies. Thus, the ER was represented by 4 arcs of knots (or segments), as 
suggested in Figure \ref{fig:segments}. Along each of the 4 arcs, 
the knots are assumed to be of uniform brightness, but the 
relative brightnesses of each arc are independent. All 7 of these 
components are convolved with an empirical PSF (for each mosaic) 
that is taken from a nearby isolated bright star in the image.
The convolved components are shown with equal normalization 
in Figure \ref{fig:segments}, to show that they do represent spatially
distinct components even after convolution. 

The model does not fit, or include any adjustment for, 
the expansion of the ER over time. The expansion rate at X-ray 
wavelengths from \cite{Frank:2016} translates to $0.02''$ per 1000 
days, or about $0.09''$ over the span of the three intervals being 
modeled. \cite{Larsson:2019} report an expansion rate almost 
3 times slower from examination of B and R band {\it Hubble} images.

A linear regression is used to solve for the amplitudes of each 
of the model components. The formal uncertainties on the amplitudes
are $\lesssim 2\%$. 
The resulting models are compared to 
the actual images in Figure \ref{fig:epochs}. The residuals 
in the difference between the data and the models do not 
suggest there is any additional component missing from the model.
Specifically, there is no indication that a central component 
representing the slower ejecta is needed.

\begin{figure}[t]
\begin{center}
\includegraphics[width=3.25in]{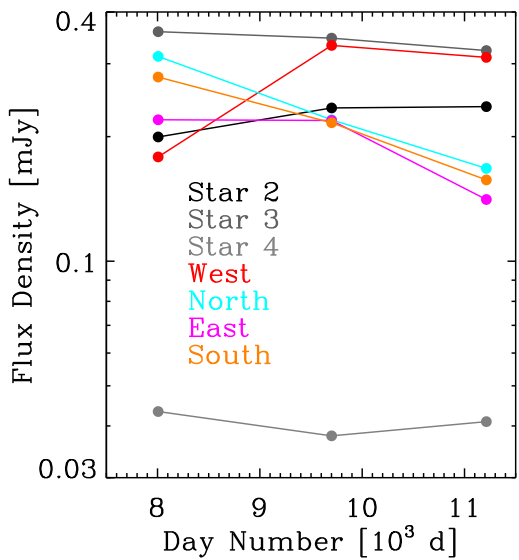}
\caption{The 3.6 $\mu$m light curves of the four ER segments 
and the 3 surrounding stars, derived from the decomposition of 
the $0.4''$ mosaics at 3 epochs. Formal uncertainties on the 
flux densities are $\lesssim 2\%$.
\label{fig:decomp_light_curves}}
\end{center}
\end{figure}

The brightnesses of the model components at early, middle, and late
epochs are plotted in Figure \ref{fig:decomp_light_curves}. The 
brightnesses are normalized such that the sum of the 4 ER components
is well matched by the integrated 3.6~$\mu$m light 
curve (Figure \ref{fig:light_curve}). 
There are moderately strong inverse correlations 
between adjacent ER segments with 
$0.34 < \sigma_{ij}^2/\sigma_i\sigma_j < 0.44$, and positive correlations
between opposing segments with $0.10 < \sigma_{ij}^2/\sigma_i\sigma_j < 0.18$.
There is also an inverse correlation between the east ER segment 
and Star 3: $\sigma_{ij}^2/\sigma_i\sigma_j = 0.22$.
These results indicate that 
the west segment of the ER has brightened substantially during 
the span of observations, while the other segments have been fading
throughout the observations. This supports and quantifies 
the apparent asymmetries in the images in Figure~\ref{fig:epochs}.

The mapping and decomposition procedures were also applied to the 
individual epochs of targeted observation without averaging into 
broader intervals. In this case, $\sim25\%$ of the $0.4''$ pixels 
in the images contain no data, and the decomposition is applied giving
no weight to these pixels. The results are, as expected, much noisier than
those illustrated above, but the same trends do emerge when similar averaging
is applied after, rather than before, the decomposition.

\section{Discussion} \label{sec:discussion}

Across all IR wavelengths, the light curves are 
reasonably well fitted by the 
model of Equation (1) with similar parameters (Figure \ref{fig:model}). 
This empirical model is similar in form 
to those developed by \cite{Lundqvist:1991} and 
\cite{Dwek:1992b} to explain the prompt line emission from the ER. 
In those models, the light travel time across the ER 
\citep[326~d for an ER with radius $R =239$~lt-days and 
inclination $i=43\arcdeg$][]{Dwek:1992b} is explicitly modeled.
Here, the light travel delays can be assumed to be a contributing 
factor to the width of the gaussian component of our model. 
Irregularities in the radius and density distribution of the ER
and asymmetries in the radius (or velocity) of the blast wave
would be other contributing factors.

The X-ray light curve is 
also well fitted by this model, although the date, $d_0$, of the peak of 
the gaussian function is somewhat later, and the exponential time scale,
$\tau_d$ is significantly longer than at the IR wavelengths. Both these
differences act to shift the peak of the X-ray light curve to a later date.
In a very simple interpretation, and assuming that the IR emission is 
from dust that is collisionally heated by the X-ray emitting 
gas \citep{Dwek:1986,Bocchio:2013}, the
shorter exponential time scale for the IR emission compared 
to that of the X-ray emission, may suggest
that dust is being destroyed faster than the gas is cooling.
However, additional factors such as the development of a reverse shock, 
and variations in the density (and consequent post-shock temperature) 
of the CSM may also contribute to the differences in 
$d_0$ and $\tau_{\rm d}$ between the IR and X-ray light curves.
Systematic differences in the IR and X-ray trends that may be indicative 
of dust destruction had been much more difficult to identify prior 
to about Day 8000 while both light curves 
were still rising \citep[e.g.][]{Dwek:2008, Dwek:2010}.

As shown in Figure \ref{fig:model}, we also find that the model of
Equation (1) provides a good fit to the evolution of the B, R, and F502N band optical 
emission reported by \cite{Larsson:2019} if restricted to dates 
after Day 5500 when the interaction with the ER is well developed.
\cite{Larsson:2019} note that the F502N band (dominated by 
[\ion{O}{3}] line emission) rises and fades more rapidly than the 
B and R band emission. We find these differences reflected more by 
the $\tau_{\rm d}$ decay rate of the model's exponential term, 
than the $\sigma_{\rm d}$ width of the gaussian term. The model's 
parameters for the optical emission are similar to the parameters
for the IR emission. Specifically, the decay time scales for the 
R and B band emission are $\lesssim30\%$ longer than $\tau_{\rm d}$ at 
4.5 $\mu$m, while the decay time scale for the X-ray emission is $>2$ 
times longer. If the dust is associated with the denser 
gas clumps that dominate the optical emission rather than the X-ray 
emitting gas, then there is little evidence from the total ER light 
curves to indicate ongoing dust destruction.
More detailed analysis about the origin and evolution of the dust 
and the infrared emission from the ER will be addressed in a 
forthcoming paper \citep{Dwek:2020}.

The deconvolution and decomposition provide conclusive evidence
that the 3.6 and 4.5~$\mu$m emission is dominated by the ER and 
not the ejecta. The decomposition also clearly shows that
at 3.6~$\mu$m the west side of the ER is brightening 
relative to the rest of the ER. The behavior is consistent 
with that observed in the better spatially-resolved and better 
temporally-sampled X-ray and optical results obtained by \cite{Frank:2016}
and \cite{Larsson:2019}. Our first time interval, Days 7156 -- 8856,
spans the time when both the optical and X-ray emission were transitioning 
from a brighter eastern side to a brighter western side. During our
latter two epochs, we find the west half of the ER is brighter at 
3.6~$\mu$m by similar proportions as at X-ray and optical wavelengths.

The 3.6 $\mu$m decomposition results imply that 
the total stellar flux density to be subtracted from 
the aperture photometry should be 0.61 mJy, rather than
the 0.41 mJy that had been used previously
(in Table \ref{tab:fluxes}, Figure \ref{fig:light_curve},
and previous papers).
Use of the larger stellar contribution would steepen the 
decline of the intensity, yielding $\tau_{\rm d} = 5628$ for 
the model shown in Figure \ref{fig:model}. 
This is closer to, but
still larger than, $\tau_{\rm d} = 5381$ derived at 4.5 $\mu$m.
Any adjustments to the assumed stellar flux densities at longer 
wavelengths would have less importance because the SN 
is much redder than the stellar sources.
\cite{Walborn:1993} identified Star 3 as a Be star, and observed
it to fade in the $J$, $H$, and $K$ bands by about one magnitude 
over an interval of $\sim1000$ days. Thus it is plausible that 
the star's brightness during the period of IRAC observations 
could be brighter (and less constant) than expected.

\section{Summary} \label{sec:summary}

We have examined the complete record of {\it Spitzer} IRAC
observations of SN 1987A which span the period from roughly 
6000 to 12000 d after the SN explosion. These data include 3.6 and 4.5~$\mu$m 
photometry as the supernova's blast wave has run into and through
the pre-existing circumstellar equatorial ring (ER). We find that 
the mid-IR light curves of the encounter can be well fitted by a model
that is the convolution of a gaussian function and an exponential
decay. The model is a good fit to the soft X-ray and optical light curves as well. 
With application of deconvolution procedures we can see that
the spatial structure of the 3.6 and 4.5~$\mu$m emission 
matches the equatorial ring rather than the inner ejecta.
By modeling the high-resolution maps of the emission at different
epochs, we find that the 3.6~$\mu$m emission has changed from 
being relatively uniform around the ER to being brighter on
the western side. 

The true nature of the emission at 3.6 and 4.5~$\mu$m should 
be made clear with the {\it James Webb Space Telescope} 
\citep[{\it JWST;}][]{Gardner:2006}, which will have vastly
improved angular resolution for distinguishing the true 
morphology of the emission and distinguishing the CSM from 
the ejecta and from nearby stars. More importantly, 
{\it JWST} will provide high resolution spectral data 
across this wavelength regime which will clearly 
identify line and continuum components, thus revealing 
the true physical source of the emission.

The {\it Spitzer} IRAC observations serendipitously 
also revealed a strong slow variable object in the same 
wide field of view as SN 1987A (see Appendix). An optical spectrum obtained 
for this source indicates that it is a previously 
unreported Be star.\\

%\acknowledgements
We thank K. Frank for providing the additional epochs of X-ray data
that are presented here, A. Kashlinsky for pointing out the possibility 
of the variable source being a microlensing event, 
and E. Pompei for subsequently obtaining 
the spectrum showing the source is a Be star.
We thank the referee for comments that led to improved clarity 
and content of the work presented here.
This work is based on observations made with the 
{\it Spitzer Space Telescope}, which is operated by the Jet Propulsion Laboratory, 
California Institute of Technology under a contract with NASA. Support for 
this work was provided by NASA. This research has made use of NASA's Astrophysics 
Data System Bibliographic Services. RDG was supported by NASA and the United States Air Force.
CEW was supported by NASA.

\facilities{{\it Spitzer} (IRAC, MIPS), {\it Chandra}}, {\it Hubble}

\software{IDLASTRO \citep{Landsman:1995}}

\section*{Appendix}
%\appendix

\section*{A second strongly variable source}

\begin{figure*}[t]
\begin{center}
\includegraphics[height=2.in]{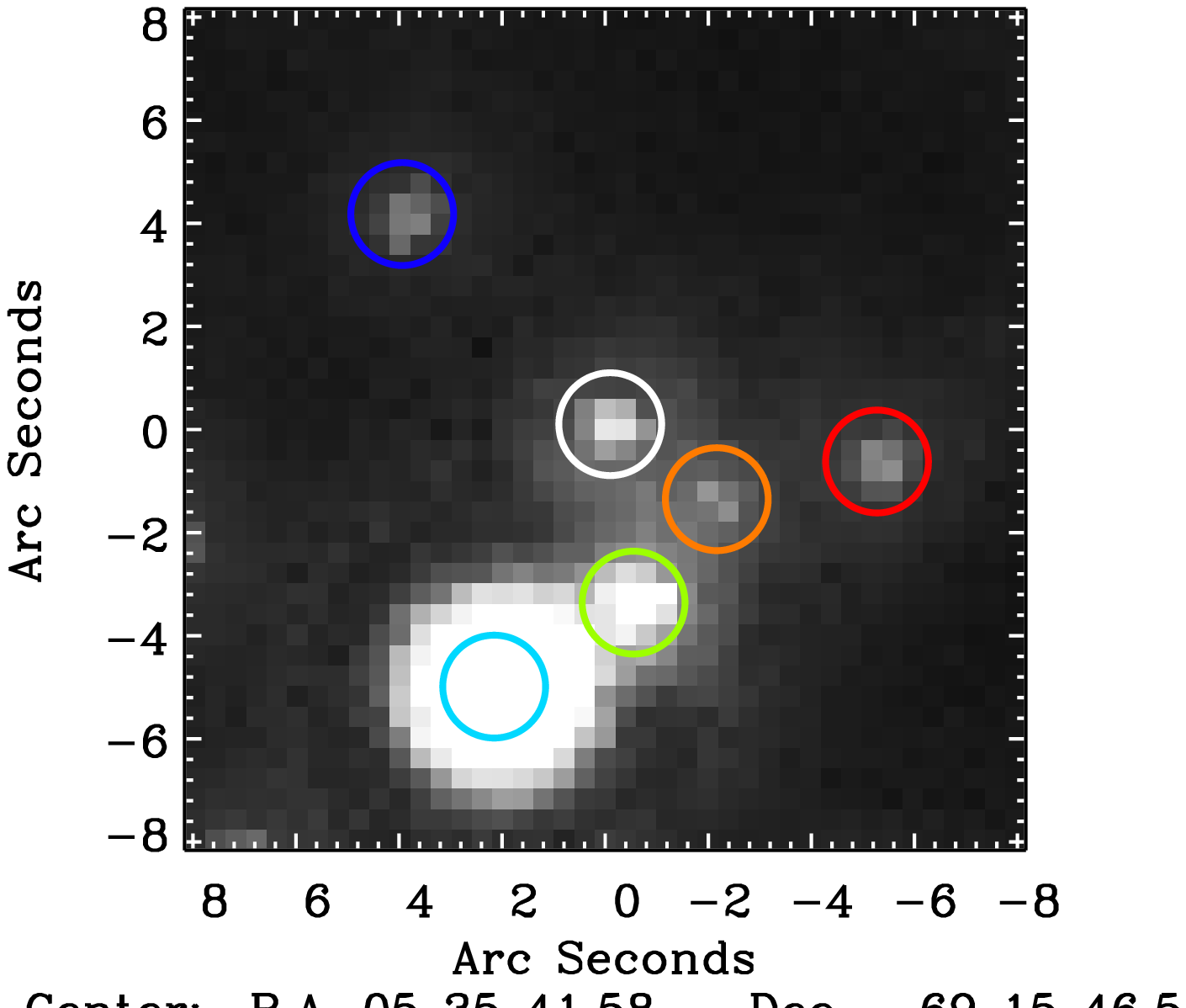}
\includegraphics[height=2.in]{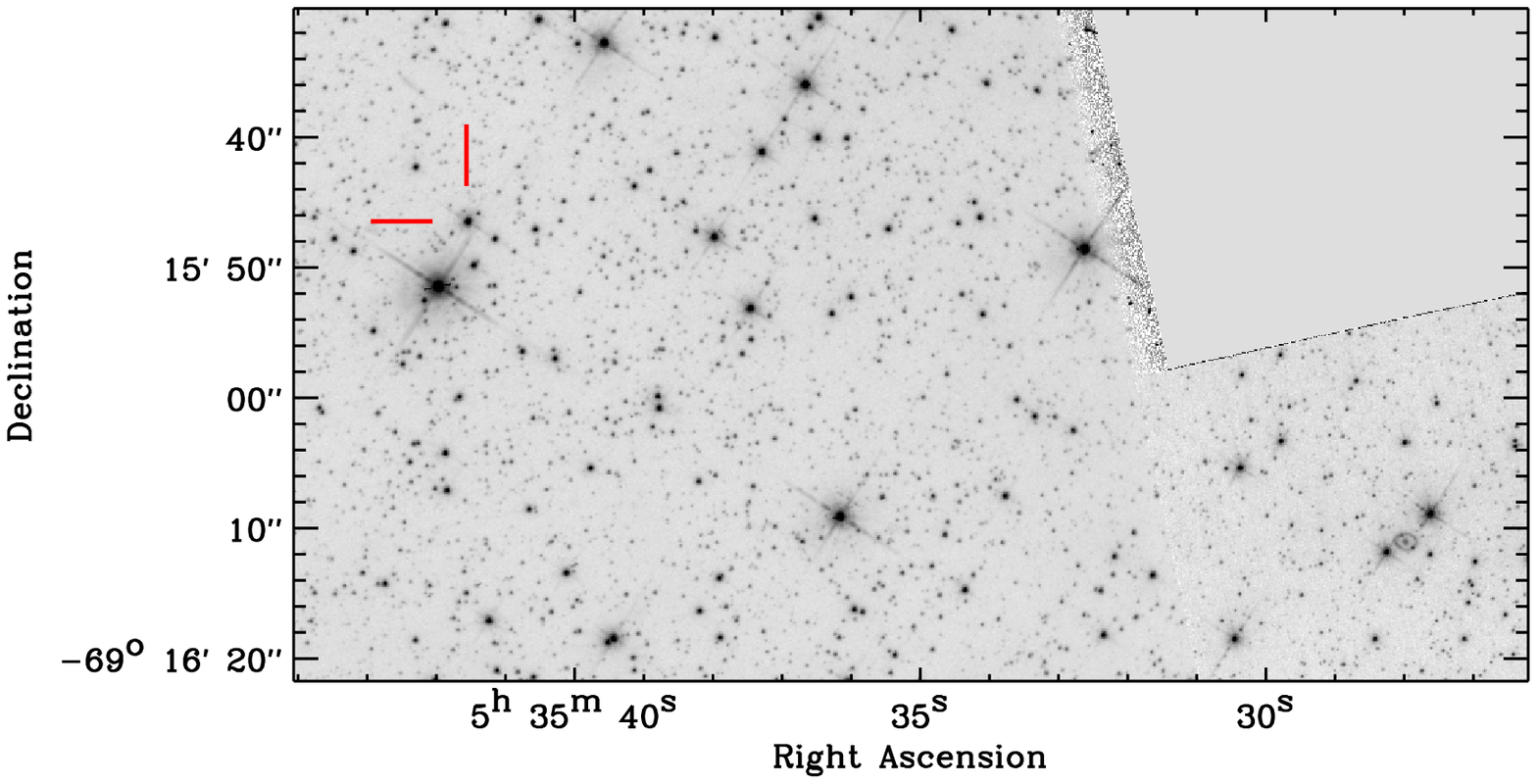}
\end{center}
\caption{(left) 3.6 $\mu$m image of the variable source (white circle) and surrounding stars that 
were simultaneously fitted (colored circles). (right) Location of the variable source (red lines) 
relative to SN 1987A (at lower right) in a WFPC2 F814W image from {\it HST}.
\label{fig:image_var}}
\end{figure*}

\begin{figure*}[t]
\includegraphics[width=4.8in]{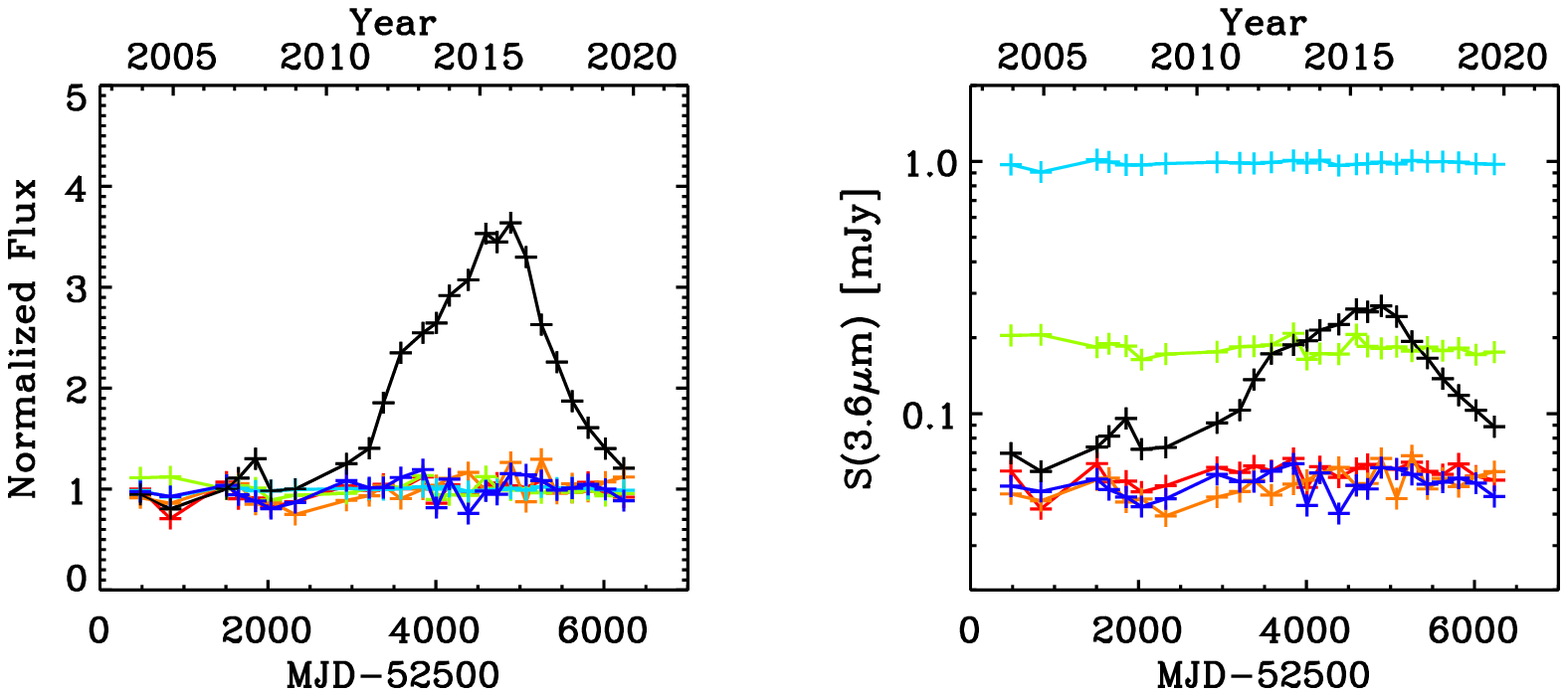}\\
\includegraphics[width=4.8in]{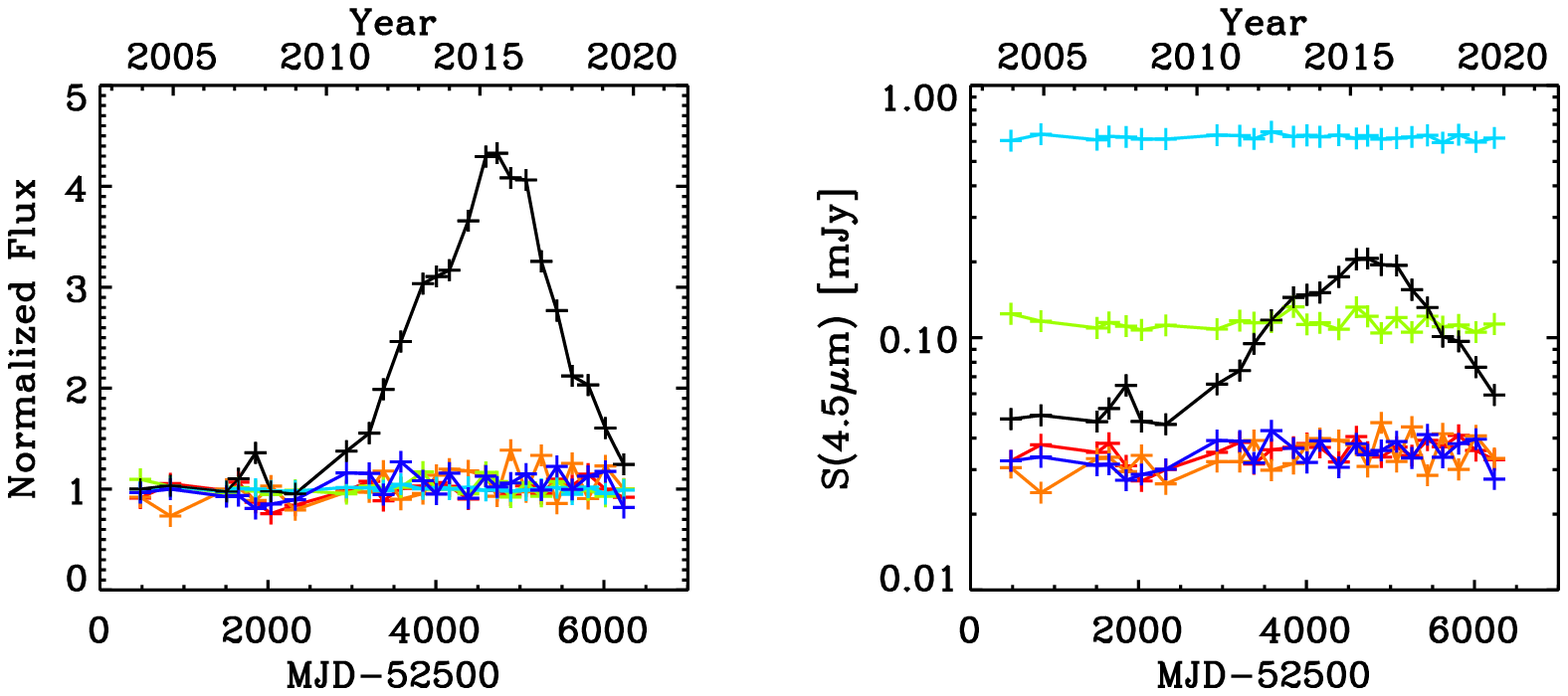}
\includegraphics[width=2.4in]{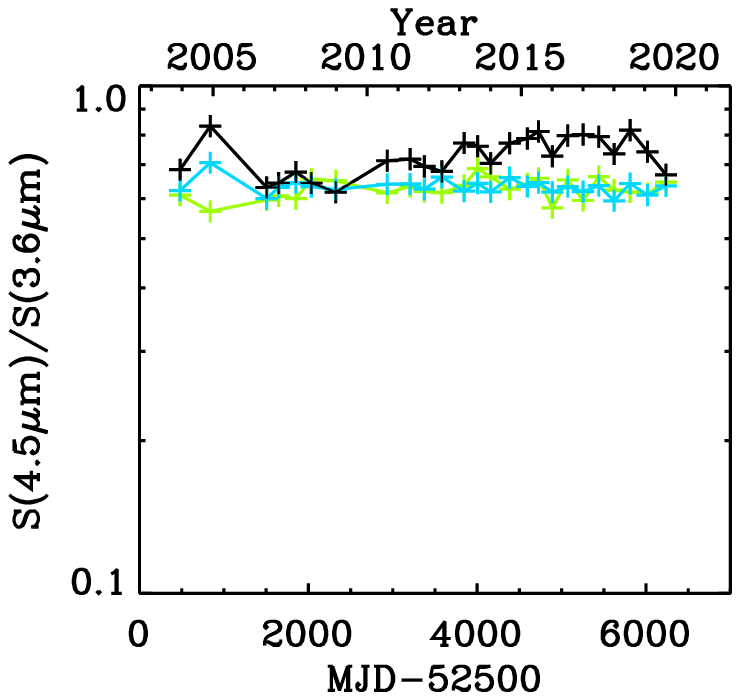}
\caption{3.6 $\mu$m (top row) and 4.5 $\mu$m (bottom row) light curves of 
the sources in Figure \ref{fig:image_var}.
In the left column all the light curves are normalized to the respective 
flux densities at early times. 
In the middle column, the absolute flux densities are shown. 
The symbol colors of the stable stars match those 
used in Figure \ref{fig:image_var} (left).
The panel at lower right shows that color of the variable source steadily 
changes with respect to the stable colors of the two brighter sources.
\label{fig:light_curve_var}}
\end{figure*}

\begin{figure}[t]
\begin{center}
\includegraphics[width=3.25in]{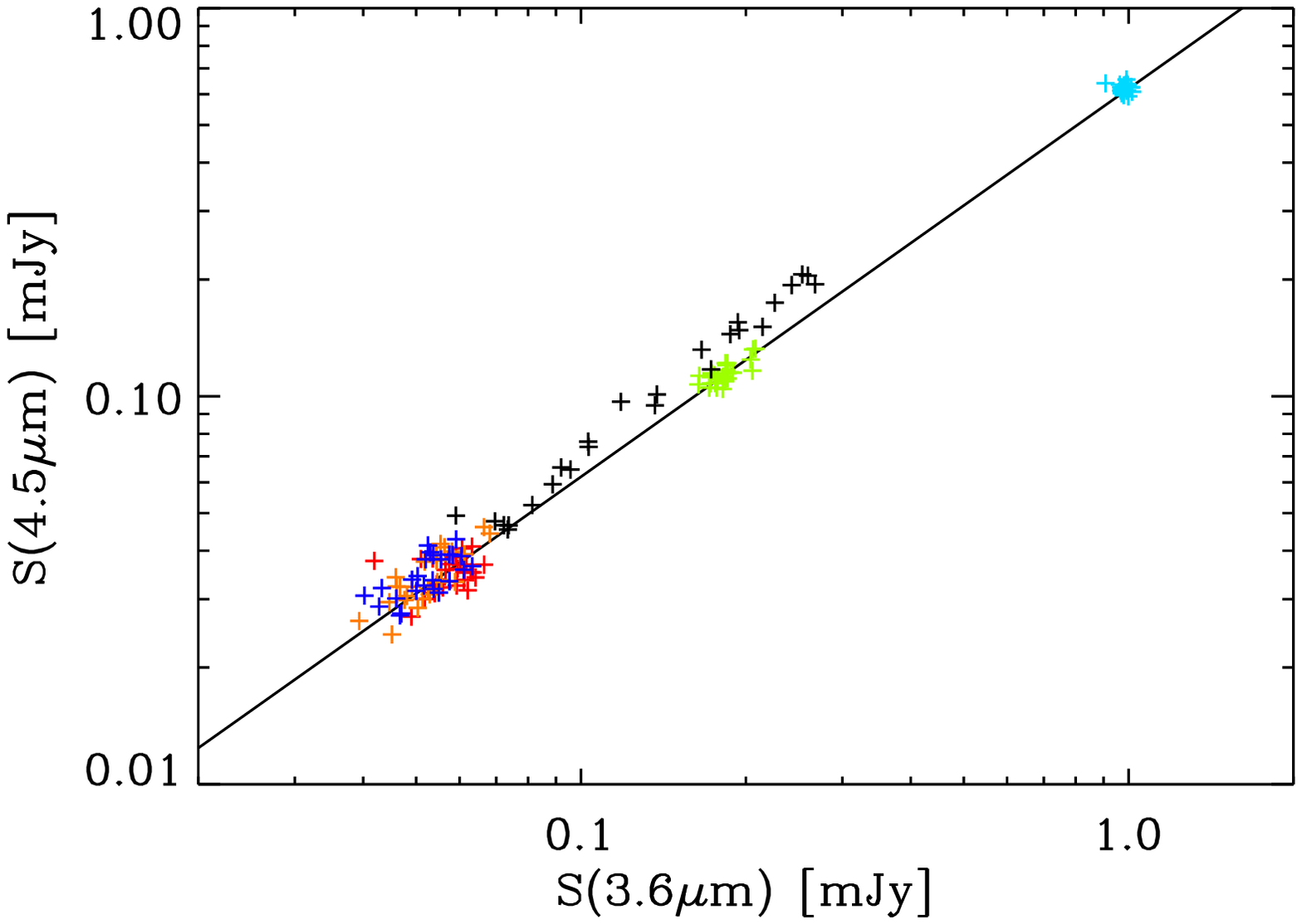}
\end{center}
\caption{The correlation between the 3.6 and 4.5 $\mu$m fluxes for the 
sources in Figure \ref{fig:image_var}. The field sources (colored symbols) 
exhibit a fixed color that is well matched to the Rayleigh-Jeans tail 
of a blackbody spectrum (trend indicated by the black line). 
The variable source (black symbols) is close to the blackbody color when faint, but
systematically reddens as it brightens.
\label{fig:color_var}}
\end{figure}

{\it Spitzer} has revealed a strongly variable point 
source\footnote{$(\alpha,\delta) = (83.923180,-69.262906)$,
\object{{\it Spitzer} SSTISAGEMC J053541.50-691546.5}, 
{\it Gaia} DR2 4657667839561179136, 
{\it Hubble} Source Catalog V3 matchID 76408986} 
that lies just $76''$ (PA = $71\arcdeg$) east of SN 1987A,
and was thus serendipitously monitored by each of the SN 1987A observations.
A 3.6 $\mu$m image of the variable and nearby field stars is shown in 
Figure \ref{fig:image_var}.
Aperture photometry does not yield stable photometry for this source because it 
is in a crowded field and it is much fainter than SN 1987A. 
Relative photometry of the variable source and field stars was performed 
by simultaneous fits of fixed-width gaussian beams to each source,
with locations fixed by the source positions in the Hubble Source Catalog
\citep[HSC,][]{Whitmore:2016}.
An initial fit and cross correlation are made to determine any fractional 
pixel shift in the nominal astrometry. After a positional adjustment,
a second fit determines the relative fluxes of each of the sources. 
The absolute flux densities are set by normalizing the mean intensity of the
brightest star (cyan symbols in 
Figures \ref{fig:image_var}, \ref{fig:light_curve_var}, and 
\ref{fig:color_var}) to the results
given in the Surveying the Agents of Galaxy Evolution (SAGE) 
Catalog \citep{Meixner:2006}.
The source flux densities are shown in Figure \ref{fig:light_curve_var}.
The light curves demonstrate the good stability of the field stars 
during a period where the variable source brightens by a factor of $\sim4$, 
and almost returns to its original brightness over a span of $\sim3000$ days. 
The 5.8 and 8 $\mu$m observations only exist prior to the main outburst.
The variable source is not detected at these wavelengths, except for a 
roughly one year interval centered at modified julian date (MJD) 
MJD-52500 = 1851 (= 2007 Sep 8) when the source does 
appear at 5.8 $\mu$m. This matches the much smaller outburst 
that occurs simultaneously at 3.6 and 4.5 $\mu$m at this time. 
The serendipitous monitoring with {\it Hubble} 
[Wide Field and Planetary Camera 2 (WFPC2) and Wide Field Camera 3 (WFC3) 
observations collected in the HSC] mostly misses the strong outburst period,
as SN 1987A was usually observed in subarray modes that used insufficiently large
fields of view to capture this source. The few observations that do exist 
suggest that the outburst was fainter or absent at shorter wavelengths.
The source is not listed in the Hubble Catalog of Variables
\citep[HCV,][]{Bonanos:2019}, but is flagged as variable in the 
Visible and Infrared Survey Telescope for Astronomy (VISTA) 
Magellanic Clouds survey \citep[VMC DR4,][]{cioni:2011}.
Variability information is listed as ``NOT$\_$AVAILABLE'' in the Gaia DR2 
catalog \citep{Gaia-Collaboration:2018, Holl:2018}.

A possible explanation for this source is that is it a classical Be star.
The brightness and colors of the star are generally consistent with a main 
sequence B star in the LMC. More specifically, the ultraviolet -- mid-IR SED 
is very similar to those of classical Be stars reported by 
\cite{Gehrz:1974} or \cite{Bonanos:2009} for example,
with excess 3.6 and 4.5 $\mu$m emission attributable to free-free emission.
A change in the circumstellar free-free emission could occur with 
relatively little impact on the brightness at shorter wavelengths.
The outburst seems rather large and long for typical Be star behavior,
but may represent an extreme example. Classical Be star outbursts
usually rise faster than they decline \citep{Rivinius:2013,Labadie-Bartz:2017}.

On 2018 Nov 26 (MJD = 58448.2), E. Pompei 
(private communication) used the European
Southern Observatory (ESO) Faint 
Object Spectrograph and Camera (EFOSC) on the ESO New Technology Telescope (NTT) to 
obtain a 3600-9200\AA\ spectrum (720 s exposure, $R\sim 500$) of the source. 
The spectrum 
confirms that this is a Be star, showing H$\alpha$ in emission. 
H$\beta$ appears as a weak absorption line,
and higher Balmer lines are clearly seen in absorption. Overall, the 
spectrum closely resembles that of the B3Ve star 
HD 191610 \citep{Valdes:2004}\footnote{ \url{https://www.noao.edu/cflib/} 
or see\\ \url{https://www.cfa.harvard.edu/~pberlind/atlas/htmls/bstars.html}}. 

The spectrum firmly rules out the alternate possibility
that the source is an active galactic nucleus or quasar. 
Archival {\it Hubble} images show no indication 
of diffuse emission (i.e. a host galaxy) around the variable point source, 
although there appears to be a background 
galaxy located $12''$ to the northeast.

A third possibility we considered is that this 
is a gravitational microlensing event. 
The long duration would imply an exceptionally 
massive lens, or an exceptionally small transverse velocity. 
However, the asymmetric light curve (and the smaller earlier 
brightening) would necessitate complex lens and/or source structures. 
A larger difficulty with this explanation is that the magnification 
should be independent of wavelength, while the observations
indicate a distinct reddening during the main outburst 
(Figures~\ref{fig:light_curve_var} and \ref{fig:color_var}).

\bibliographystyle{aasjournal}
\bibliography{SN1987A_final_irac}

\end{document}